\documentclass[fleqn,usenatbib]{mnras}

\usepackage{newtxtext,newtxmath}

\usepackage[T1]{fontenc}
\usepackage{ae,aecompl}

\usepackage{graphicx}	
\usepackage{amsmath}	
\usepackage{amssymb}	
\usepackage[usenames, dvipsnames]{color}

\usepackage{float}

\usepackage{multicol}
\usepackage{amsmath}	
\usepackage{amssymb}	
\usepackage{caption}
\usepackage{enumerate}
\usepackage{multirow}

\newcommand{\nmx}{$\nu_{\rm{max}}$ }
\newcommand{\dnu}{$\Delta\nu$ }
\def\code#1{\texttt{#1}}

\title[K2 Galactic Caps Project]{The K2 Galactic Caps Project - Going Beyond the \textit{Kepler} Field and Ageing the Galactic Disc}

\author[B. M. Rendle et al.]
{B.M. Rendle$^{1,2}$\thanks{E-mail: bm.rendle8@gmail.com}, 
A. Miglio$^{1,2}$, 
C. Chiappini$^{3}$, 
M. Valentini$^{3}$,
G.R. Davies$^{1,2}$,
\newauthor B. Mosser$^{4}$,
Y. Elsworth$^{1,2}$,
R.A. Garc\'ia$^{5,6}$,
S. Mathur$^{7,8,9}$,
P. Jofr\'e$^{10}$,
C.C. Worley$^{11}$,
\newauthor L. Casagrande$^{12}$,
L. Girardi$^{13}$,
M.N. Lund$^{2,1}$,
D.K. Feuillet$^{14}$,
A. Gavel$^{15}$,
L. Magrini$^{16}$,
\newauthor S. Khan$^{1,2}$,
T.S. Rodrigues$^{13}$,
J.A. Johnson$^{17,18}$,
K. Cunha$^{19,20}$,
R. L. Lane$^{21,22}$,
\newauthor C. Nitschelm$^{23}$, W.J. Chaplin$^{1,2}$
\\
$^{1}$School of Physics and Astronomy, University of Birmingham, Birmingham, B15 2TT, UK\\
$^{2}$Stellar Astrophysics Centre, Department of Physics and Astronomy, Aarhus University, Ny Munkegade 120, DK-8000 Aarhus C, Denmark\\
$^{3}$Leibniz-Institut f\"{u}r Astrophysik Potsdam (AIP), An der Sternwarte 16, 14482, Potsdam, Germany\\
$^{4}$LESIA, Observatoire de Paris, PSL Research University, CNRS, Sorbonne Universit\'e, Universit\'e Paris Diderot, 92195 Meudon, France\\
$^{5}$IRFU, CEA, Universit\'e Paris-Saclay, 91191 Gif-sur-Yvette, France\\
$^{6}$AIM, CEA, CNRS, Universit\'e Paris-Saclay, Universit\'e Paris Diderot, Sorbonne Paris Cit\'e, 91191 Gif-sur-Yvette, France\\
$^{7}$Instituto de Astrof\'isica de Canarias, 38200 La Laguna, Tenerife, Spain\\
$^{8}$Universidad de La Laguna, Dpto. de Astrof\'isica, 38205 La Laguna, Tenerife, Spain\\
$^{9}$Space Science Institute, 4750 Walnut Street Suite 205, Boulder, CO 80301, USA\\
$^{10}$N\'ucleo de Astronom\'ia, Facultad de Ingenier \'ia y Ciencias, Universidad Diego Portales, Av.  Ej\'ercito 441, Santiago, Chile\\
$^{11}$Institute of Astronomy, University of Cambridge, Madingley Rise, Cambridge CB3 0HA\\
$^{12}$Research School of Astronomy and Astrophysics, Mount Stromlo Observatory, The Australian National University, ACT 2611, Australia\\
$^{13}$Osservatorio Astronomico di Padova - INAF, Vicolo dell'Osservatorio 5, I-35122 Padova, Italy\\
$^{14}$Max-Planck-Institut f\"{u}r Astronomie, K\"{o}nigstuhl 17, D-69117 Heidelberg, Germany\\
$^{15}$Department of Physics and Astronomy, Uppsala University, Box 516, 75120 Uppsala, Sweden\\ 
$^{16}$INAF  -  Osservatorio  Astrofisico  di  Arcetri,  Largo  E.  Fermi,  5,  I-50125 Firenze, Italy\\
$^{17}$Department of Astronomy, Ohio State University, 140 W 18th Ave, Columbus, OH 43210, USA\\
$^{18}$Center for Cosmology and AstroParticle Physics, 191 West Woodruff Avenue, Ohio State University, Columbus, OH, 43210, USA\\
$^{19}$University of Arizona, Tucson, AZ 85719, USA\\
$^{20}$Observat\'orio Nacional, S\~ao Crist\'ov\~ao, Rio de Janeiro, Brazil\\
$^{21}$Instituto de Astrof{\'i}sica, Pontificia Universidad Cat{\'o}lica de Chile, Av. Vicuna Mackenna 4860, 782-0436 Macul, Santiago, Chile\\
$^{22}$Millennium Institute of Astrophysics, Av. Vicu\~na Mackenna 4860, 782-0436 Macul, Santiago, Chile\\
$^{23}$Centro de Astronom{\'i}a (CITEVA), Universidad de Antofagasta, Avenida Angamos 601, Antofagasta 1270300, Chile\\
}
\date{Accepted XXX. Received YYY; in original form ZZZ}

\pubyear{2019}

\begin{document}
\label{firstpage}
\pagerange{\pageref{firstpage}--\pageref{lastpage}}
\maketitle

\begin{abstract}
\vspace{-0.05cm}
Analyses of data from spectroscopic and astrometric surveys have led to conflicting results concerning the vertical characteristics of the Milky Way. Ages are often used to provide clarity, but typical uncertainties of $>$ 40\,\% from photometry restrict the validity of the inferences made. Using the \textit{Kepler} APOKASC sample for context, we explore the global population trends of two K2 campaign fields (3 and 6), which extend further vertically out of the Galactic plane than APOKASC. We analyse the properties of red giant stars utilising three asteroseismic data analysis methods to cross-check and validate detections. The Bayesian inference tool PARAM is used to determine the stellar masses, radii and ages. Evidence of a pronounced red giant branch bump and an [$\alpha$/Fe] dependence on the position of the red clump is observed from the K2 fields radii distribution. Two peaks in the age distribution centred at $\sim$5 and and $\sim$12 Gyr are found using a sample with $\sigma_{\rm{age}}$ $<$ 35\,\%. In comparison with \textit{Kepler}, we find the older peak to be more prominent for K2. This age bimodality is also observed based on a chemical selection of low- ($\leq$ 0.1) and high- ($>$ 0.1) [$\alpha$/Fe] stars. As a function of vertical distance from the Galactic mid-plane ($|Z|$), the age distribution shows a transition from a young to old stellar population with increasing $|Z|$ for the K2 fields. Further coverage of campaign targets with high resolution spectroscopy is required to increase the yield of precise ages achievable with asteroseismology.

\end{abstract}


\section{Introduction}

Understanding and classifying the fundamental properties and formation mechanisms of galaxies is a cornerstone of characterising the evolutionary processes of both galactic and large scale extra-galactic structures. Galactic archaeology is a rapidly expanding field, using fossil remnants within the Milky Way to understand its formation history. The objective of the field is to understand the mechanisms of formation and structure of the Galaxy through the study of the collective properties of stellar populations. Accessing and correctly interpreting this information is key when wanting to understand Galactic evolution, especially during its earliest phases. High red-shift disc galaxies appear to undergo the most significant formation changes between 12 and 8 Gyr ago at $z \approx 2$ (e.g. \citealt{2014ARA&A..52..415M}). The expected bulge, halo and disc structures are typically formed during this time, with only thin disc formation steadily continuing to the present. This has been predicted by multiple theoretical models (e.g. \citealt{2003ApJ...591..499A, 2013ApJ...773...43B, 2009ApJ...707L...1B,  2004ApJ...612..894B, 2009IAUS..254..445G, 2013ApJ...772...36G, 1983A&A...120..165J, 2016AN....337..976K, 1998Natur.392..253N, 2003ApJ...596...47S, 1994A&A...281L..97S}) and also appears to be true for the Milky Way (\citealt{1997ApJ...477..765C,2009IAUS..254..191C}; \citealt{ 2015A&A...580A.126K, 2013A&A...558A...9M,2014A&A...572A..92M, 2015A&A...578A..87S}). Current studies imply that the formation of the thick disc in the Milky Way started at z $\sim$ 3.5 (12 Gyr), whilst thin disc formation began at $z \sim$ 1.5 (8 Gyr) (e.g. \citealt{2014A&A...562A..71B,  2014A&A...565A..89B,  2011MNRAS.414.2893F,  2013A&A...560A.109H,2014A&A...569A..13R,2018Natur.563...85H}).

There are many unanswered questions in the formation of the Milky Way (e.g. see \citealt{2017AN....338..644M,2016AN....337..703M} for a review). One of the most fundamental questions is the characterisation of its vertical structure. It is commonly agreed that the Galaxy consists of a central bar/bulge, disc and halo components (e.g. \citealt{2016ARA&A..54..529B,2018ARA&A..56..223B}). The specific nature of each component has been subject to scrutiny, with the nature of the disc most fervently debated (see e.g. \citealt{2016AN....337..976K}). Since the results showing evidence for a multiple disc-like structure \citep{gilmore_new_1983}, astronomers have striven to fully classify these components and distinguish them chemically, dynamically and geometrically (see e.g. \citealt{2002ARA&A..40..487F,2012ApJ...751..131B,2012ApJ...755..115B,2013A&A...560A.109H,2013A&ARv..21...61R,2014IAUS..298...17B,2014A&A...564A.115A,2014A&A...567A...5R,2015ApJ...808..132H,2015ApJ...804L...9M,2016ApJ...823...30B,2017A&A...608L...1H}). Typical constraints from the literature define the discs as such: Thin Disc - scale height $\sim 300$ pc, age $\lesssim 9$ Gyr, solar-[Fe/H], solar-[$\alpha$/Fe]; Thick Disc - scale height $\sim 900$ pc, age $\gtrsim 10$ Gyr, [Fe/H] $\sim -0.7$, enhanced [$\alpha$/Fe] ($> 0.2$). Large scale spectroscopic and kinematic surveys have allowed the dissection of mono-age and mono-abundance populations (e.g. \citealt{2012ApJ...753..148B,2016ApJ...823...30B,2016ApJ...831..139M,ted_mackereth_agemetallicity_2017,Mackereth2019}), giving snapshots into different epochs of the Milky Way's past. Most studies concur on the existence of multiple stellar populations within the Galactic disc, but stress the importance of which metric is used to define the so-called thin and thick disc components respectively \citep{2012ApJ...751..131B,2012ApJ...755..115B,2015ApJ...804L...9M}, if the disc is to be classified as such. 

Any inferences to be made about Galactic structure and evolution rely heavily upon having accurate measurements of the stellar population parameters (e.g. ages, metallicities). The relevance of asteroseismology in stellar populations studies was recognised early on when the first data from CoRoT and {\it Kepler} became available \citep[see, e.g.][]{Miglio_2009, Chaplin_2011}. 
Subsequently, tests of the precision and accuracy of the asteroseismically inferred parameters  enabled quantitative studies that made use of distributions of stellar masses and wide age bins \citep{2013MNRAS.429..423M,Casagrande_2014, 2017A&A...597A..30A}.

This field has continued to mature alongside  data-analysis and modelling procedures. It is now recognised that asteroseismic constraints coupled with high resolution spectroscopy enable inferences on stellar masses, radii and ages with uncertainties of $\sim$3-10$\%$, $\sim$1-5$\%$ and $\sim$20-40$\%$ respectively (see \citealt{2016AN....337..774D}, \citealt{Mosser_2019}, where seismic yields from different observations are discussed). These uncertainties (in particular in age; see \citealt{2010ARA&A..48..581S} for a comprehensive review of determination methods) are not yet regularly achievable with spectroscopy alone, thus presenting asteroseismology as an attractive prospect for making precise parameter determinations on a large scale.

The upper limits of the asteroseismic age uncertainties are typically achievable using the masses obtained from the so-called asteroseismic scaling relations \citep{1995A&A...293...87K}. 
Stellar mass is a particularly valuable constraint in
the case of giants, since for these stars age is primarily
a function of mass. The age of low-mass red-giant stars
is largely determined by the time spent on the main
sequence, hence by the initial mass of the red giant's
progenitor ($\tau_{\rm{MS}} \propto M/L(M) \propto M_{\rm{ini}}^{-(\nu-1)}$, with $\nu = 3-5$, e.g. see  \citealt{2012sse..book.....K}).
Though only approximated relations, the masses derived usually translate into more precise ages than those achievable using photometric values and isochrone fitting.




\begin{figure}
	\includegraphics[width=0.48\textwidth]{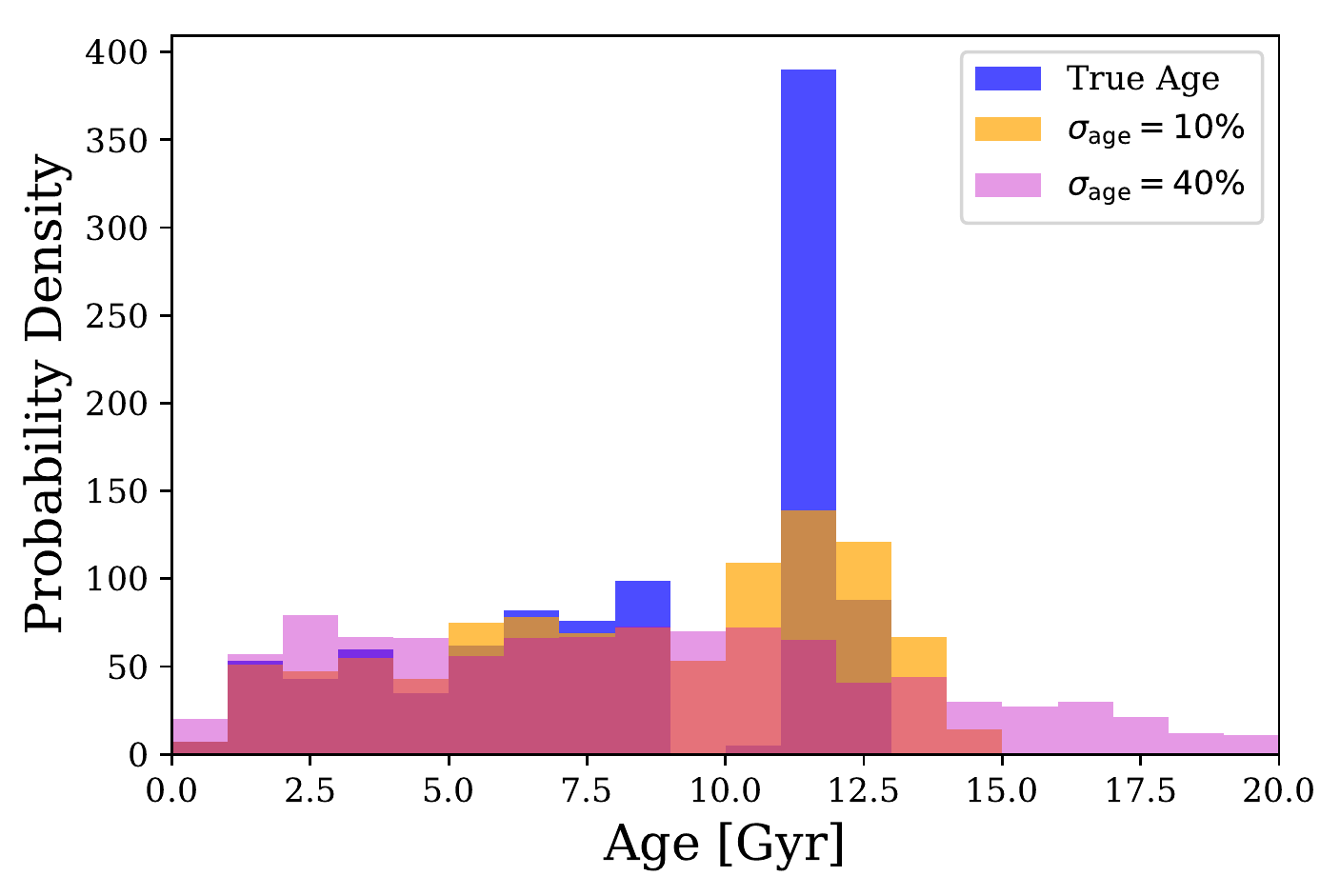}
    \caption{The normalised age distribution for a synthetic MW population (TRILEGAL) is shown (blue). This population is perturbed by age uncertainties of 10$\%$ (orange) and 40$\%$ (green) to demonstrate the necessity for high precision age determinations. It is clear that even at 10$\%$ some structural details of the population are blurred, with all structure lost when the uncertainty is 40$\%$.}
    \label{fig:Age_KDE}
\end{figure}

A 30-35$\%$ uncertainty in age is sufficient to pull out basic features of a distribution, but is not enough to conclusively interpret the true nature of the underlying population distribution. Figure \ref{fig:Age_KDE} illustrates this, showing the effect of different uncertainties on the appearance of the population distribution. Crucially, further accuracy can be achieved when one goes beyond the scaling relations, using asteroseismic grid modelling (inclusion of global asteroseismic parameters in modelling process e.g. \textsc{PARAM}, \citealt{2006A&A...458..609D, 2014MNRAS.445.2758R,2017MNRAS.467.1433R}; BASTA, \citealt{2015MNRAS.452.2127S}) or where possible, the individual acoustic modes themselves (see e.g. \citealt{2019MNRAS.484..771R} and references therein). When these techniques are implemented, it is possible to achieve the lower bounds of precision quoted. This precision greatly reduces any ambiguity surrounding the mass, radius and age distributions, allowing confidence to be given to statements regarding the state of the Milky Way at given epochs. 

Though powerful in its capabilities, asteroseismology has been relatively limited to observations of the Galactic mid-plane. CoRoT \citep{2006ESASP1306...33B,2017A&A...597A..30A} observed regions in the Galactic inner and outer disk and \textit{Kepler} \citep{Borucki977} provided exquisite data for a single field extending out of the Galactic plane. Neither mission, however, sampled sufficient fields for mapping radially and vertically the Milky Way. K2 \citep{2014PASP..126..398H} has revolutionised this, with $\sim 80$ day observations in the ecliptic plane sampling a broad range of Galactic fields to depths of several kilo-parsecs (kpc). The depth of observations and ability to detect asteroseismic signatures of extensive populations has transformed K2 into an exciting prospect for the provision of improved constraints on Galactic evolution and structure \citep{2015ApJ...809L...3S}.

The capability of asteroseismology to determine vertical stellar population trends out to and beyond $\sim1.5$ kpc has already been illustrated with the exquisite data from the CoRoT \citep{2013MNRAS.429..423M} and \textit{Kepler} \citep{2016MNRAS.455..987C,2016ApJ...827...50M,2018MNRAS.475.5487S} missions. Re-purposed as K2, asteroseismic observations towards the Galactic poles extend substantively beyond 1.5 kpc, facilitating the first detailed examination of the vertical Galactic structure with asteroseismology. Though degraded in comparison to \textit{Kepler}, the K2 data remains of high enough quality to make precise asteroseismic inferences (see \citealt{2015PASP..127.1038C,2015ApJ...809L...3S,2016MNRAS.461..760M}). Hence, using K2 campaigns 3 and 6, we present an asteroseismic analysis of the vertical disc structure of the Milky Way with the K2 Galactic Caps Project (K2 GCP). We demonstrate the increased capability of these campaign fields compared to \textit{Kepler} in determining vertical population trends and show the benefits of improved precision in age.

The paper is laid out as follows: Section \ref{data} describes the K2 campaign fields and used with the different sources of asteroseismic, spectroscopic and photometric data. Comparisons between data sources are made and effects of the selection function explored. Section \ref{method} briefly details the grid modelling tool used. Sections \ref{radii}, \ref{mass} and \ref{ages} display the key results of the work, based on the analysis of the distributions in radius, mass, and age, of the red giants observed in the two fields C3 and C6 observed by K2. Finally, section \ref{conclusion} summarises our findings and discusses the potential of future work.

\begin{table*}
	\caption{Population samples used throughout this work. Name and descriptions of the populations are provided.}
	\label{tab:cross_refs}
	l\begin{tabular}{l | l} 
		\hline
		Sample & Description\\
		\hline
        K2 & K2 sample containing parameters from the K2 EPIC catalogue.\\
        K2 Spec. & K2 cross-matched with spectroscopic surveys. Survey [Fe/H], [$\alpha$/Fe] (where applicable) and $T_{\rm{eff}}$ values used.\\
        K2 SM & K2 sample cross-matched with the SkyMapper survey. SkyMapper [Fe/H] and $T_{\rm{eff}}$ values used.\\
        APOKASC/\textit{Kepler} & \textsc{PARAM} results for the APOKASC-2 population from Miglio et al. (in prep.) .\\
        APOKASC $\alpha$-rich & $\alpha$-rich APOKASC-2 sample from Miglio et al. (in prep.).\\
        K2 $\alpha$-rich & K2 Spec. sample with [$\alpha$/Fe] $> 0.1$.\\
        K2$_{\rm{Gaia}}$ & As K2 SM sample, but with radii calculated from \textit{Gaia} parallaxes.\\
        K2$_{\rm{HQ}}$ & K2 SM sample; all stars with $\sigma_{\rm{age}} < 35\%$.\\
        K2 Spec.$_{\rm{HQ}}$ & K2 Spec. sample; all stars with $\sigma_{\rm{age}} < 35\%$.\\
		\hline        
	\end{tabular}
\end{table*}

\section{Data} \label{data}

The K2 mission provided photometric data for a range of fields located both in and out of the Galactic plane for a total of 4 years, observing 20 campaign fields (C0-19). A dedicated program for Galactic archaeology has been implemented, with observations of asteroseismic candidates in Galactic regions never previously explored with this technique on this scale. \cite{2015ApJ...809L...3S} presents the asteroseismic results for K2 campaign 1, highlighting the potential of the mission and its capabilities. Of the 20 campaign fields, nine focus on the northern and southern Galactic caps. Asteroseismic analysis of these campaign fields will improve the characterisation of the stellar populations in these directions, which in turn will assist in improving our understanding of the vertical structure.

The K2 Galactic Archaeology Project (K2 GAP, \citealt{2017ApJ...835...83S}) focuses on the observations of thousands of red giants in each K2 campaign field for the purpose of performing Galactic archaeology with potential asteroseismic targets. Red giant stars are preferentially selected over dwarfs for the K2 GAP as they are intrinsically more luminous, accommodating observations to greater distances and more detail about the Galactic structure to be probed. They also show greater oscillation amplitudes than dwarfs with frequencies well suited to the main long-cadence mode of K2. This allows for asteroseismic detections to be made for a greater sample of the observed population and consequently a more robust analysis of the population parameters.

K2 campaign fields 3 (centred at: $l = 51.1^{\circ}, b = -52.5^{\circ}$) and 6 (centred at: $l = 321.3^{\circ}, b = +49.9^{\circ}$) were selected for this work. Campaign field 3 is a south Galactic cap pointing field, whereas campaign 6 points towards the north Galactic cap.  
Red giant stars in these fields sample Galactic radii typically in the range 6-8 kpc, pointing towards the Galactic centre and observe stars up to 4kpc above and below the plane. The depth and range of these observations make these two campaign fields ideal for studies of the vertical properties of the Milky Way as both the so-called thin and thick disc populations are expected to be sufficiently sampled.

The observations of the Galactic poles are not limited to two campaign fields. Campaigns 1, 10, 14 and 17 observed the northern, and campaigns 8, 12 and 19 have observed the southern, Galactic cap. Though data is available for these fields, they are not included in this work. We limit ourselves to C3 and C6 to explore the potential of K2 to distinguish trends in the vertical Galactic structure prior to a comprehensive, multi-campaign analysis.

\begin{figure}
	\centering
    \includegraphics[width=\columnwidth]{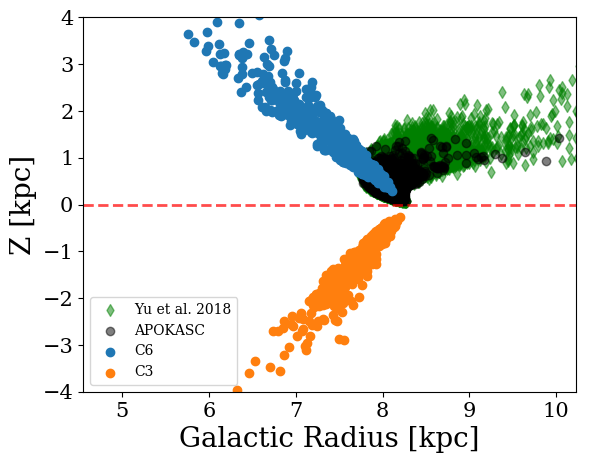}
    \caption{The distribution of stars in the APOKASC (black) and K2 campaigns 3 (orange) and 6 (blue). All $Z$ and $R_{\rm{Gal}}$ values were calculated using asteroseismic distances. The black cross shows the typical uncertainties in $Z$ and $R_{\rm{Gal}}$ of the combined K2 sample. The sample of 16000 red giants from the \textit{Kepler} survey (green diamonds, \citealt{2018ApJS..236...42Y}) shows the full range of the \textit{Kepler} field compared to the APOKASC sample.}
    \label{fig:RZ_dist}
\end{figure}

For comparative purposes, stars from the nominal \textit{Kepler} mission are included in this work. The stars were selected from the APOKASC catalogue \citep{2014ApJS..215...19P,2017AAS...22930505P,2018AAS...23145013P}. Figure \ref{fig:RZ_dist} shows the spatial distribution of the APOKASC sample compared to the respective K2 campaign fields used here. The values of $Z$ (vertical distance above the Galactic plane) and $R_{\rm{Gal}}$ (galacto-centric radius) were determined using asteroseismic distances inferred using the Bayesian inference code \textsc{PARAM} \citep{2017MNRAS.467.1433R}, with an uncertainty typically below 6\,\%. \textit{Gaia} DR2 \citep{refId0} distances are available for these stars, but the precision of asteroseismic distances in has been shown to be approximately a factor of 2 better than current \textit{Gaia} distances for the \textit{Kepler} and K2 C3 and C6 fields \citep{2019arXiv190405676K}, therefore we adopt the asteroseismic measurements here. However,
\textit{Gaia} parallaxes are used for the determination of stellar radii to investigate the target selection function (see section \ref{sel_fun}).

The extent to which the K2 fields probe vertically compared to \textit{Kepler} illustrates why this sample is suited for studies of the Galactic structure. It is expected that thick disc members dominate the stellar population beyond $|Z| \sim$1.5kpc, a region poorly sampled by \textit{Kepler} but with significant coverage by K2 across both fields. Increased coverage of stars beyond this distance is crucial for ensuring that a significant thick disc population is sampled and characterised for definitive conclusions on underlying population trends.

It is also notable from Fig. \ref{fig:RZ_dist} that the two K2 campaign fields also explore regions in the inner disc, compared to the \textit{Kepler} APOKASC population which is largely restricted to solar Galactocentric radii. Though our study focuses on the vertical properties of the field, the different pointings may be a cause of variability due to differing radial distributions. We take steps to account for this in section \ref{ages}.

Multiple subsets of targets are used in this work and for clarity they are named in Table \ref{tab:cross_refs}. The population name and a brief description of the sample are given for reference. We use a combination of asteroseismic, spectroscopic and photometric data to perform the subsequent analysis. The data used are described in turn below.

\subsection{Asteroseismology} \label{seismo}

Asteroseismic constraints were obtained using three independent asteroseismic analysis methods - BHM (Elsworth et al. in preparation), A2Z \citep{mathur_automatic_2010}, COR \citep{mosser_detecting_2009}. Multiple values for the global asteroseismic parameters \nmx and \dnu were desired to ensure that the final parameters used in the analysis were accurate, allowing for improved inputs for the grid modelling.

The same K2P$^{2}$ \citep{lund_k2p2_2015} light curves for each K2 campaign were used by each contributor. The same sample of stars was analysed by each method to extract the global asteroseismic parameters: the average large frequency separation ($\Delta\nu$) and the frequency of maximum power ($\nu_{\rm{max}}$) for each light curve. All of the methods utilise a different method to extract these global parameters. In most cases, multiple methods return a positive detection for the same star. There are also cases where a single method has registered a detection where the others have not.

\begin{figure*}
	\begin{tabular}{cc}
	\includegraphics[width=0.48\textwidth,keepaspectratio]{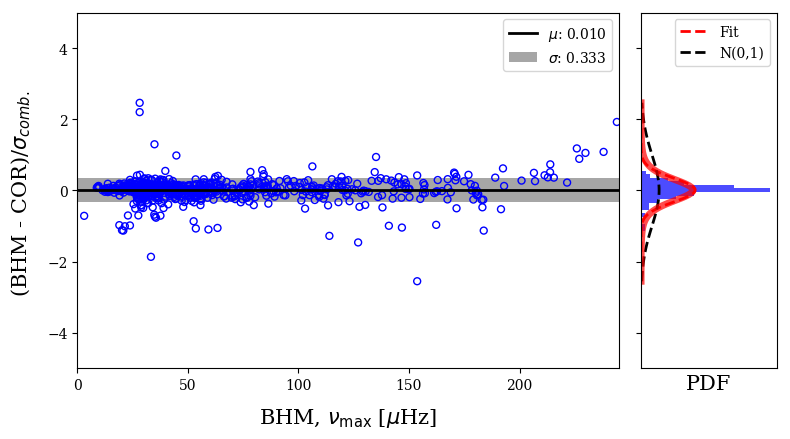} &
    \includegraphics[width=0.48\textwidth,keepaspectratio]{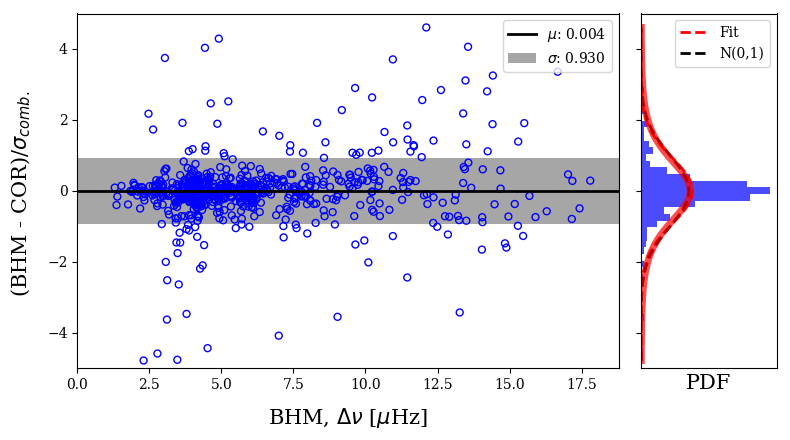}\\
    \includegraphics[width=0.48\textwidth,keepaspectratio]{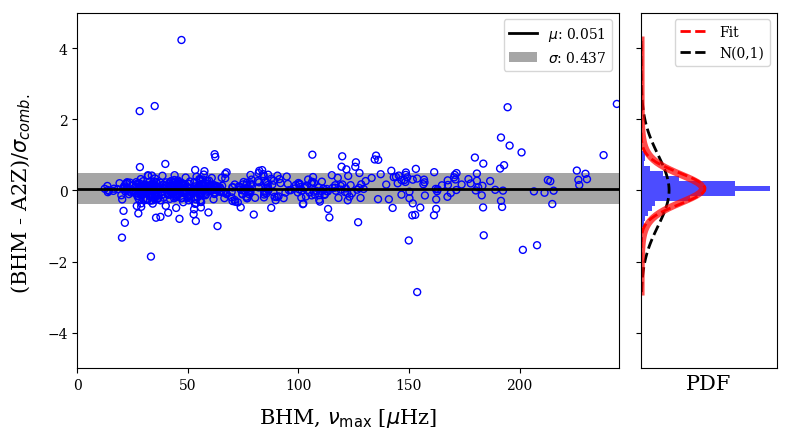} &
    \includegraphics[width=0.48\textwidth,keepaspectratio]{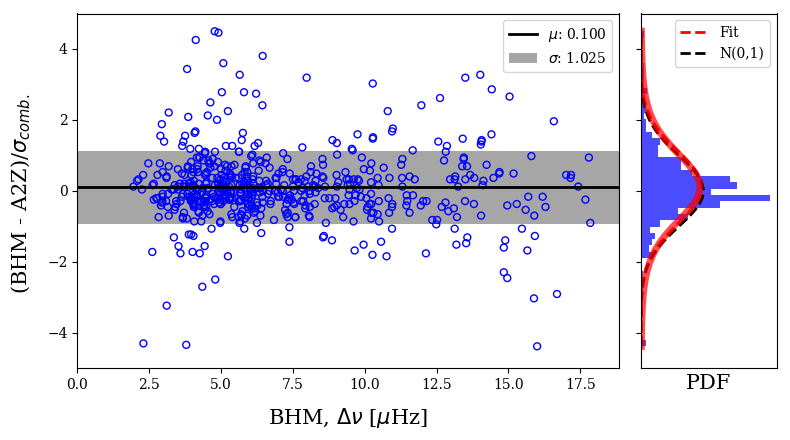}\\
    \includegraphics[width=0.48\textwidth,keepaspectratio]{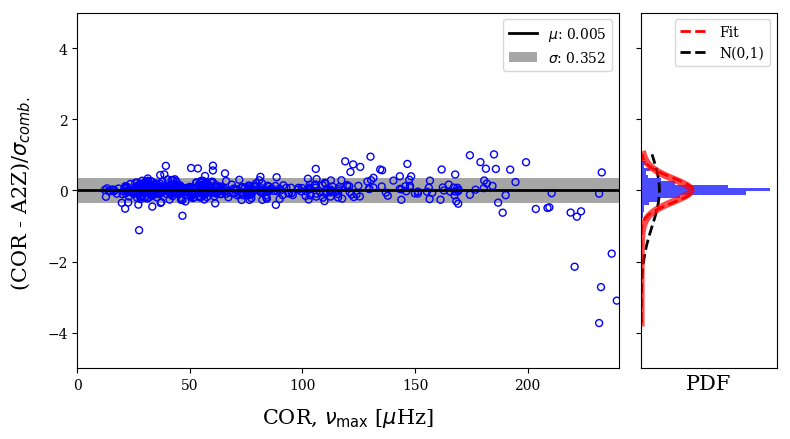} &
    \includegraphics[width=0.48\textwidth,keepaspectratio]{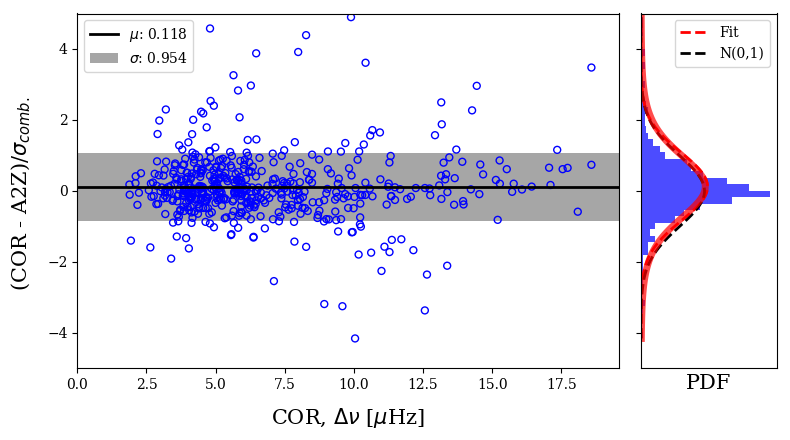}\\
    \end{tabular}
    \caption{C3 asteroseismic analysis comparisons. \textit{Left:} $\nu_{\rm{max}}$ comparisons. \textit{Right:} $\Delta\nu$ comparisons. \textit{Top:} BHM vs COR. \textit{Middle:} BHM vs A2Z. \textit{Bottom:} COR vs A2Z. Black lines show the mean ($\mu$) and the grey regions the 1$\sigma$ region of the scatter about the mean. Values are shown the legend. The Histograms show the distribution of points as a function of N$_{\sigma_{\rm{comb}}}$. Red lines show a Gaussian fit to the data using the values of $\mu$ and $\sigma$ indicated in the legend of the main panel. Black dashed lines show a N(0,1) distribution for comparison.}
    \label{fig:seismo_comp}
\end{figure*}

Having a detection from multiple methods provides an excellent opportunity to explore the consistency of different methodologies and verification of the results. Figure \ref{fig:seismo_comp} displays comparisons of \nmx and \dnu values between each of the asteroseismic methods for C3. The distribution of differences between values for crossover stars as a function of the combined uncertainty ($\sigma_{\rm{comb.}}$, mean uncertainty from cross-matches of all methods summed in quadrature) is shown. The mean ($\mu$) and standard deviation ($\sigma$) of each distribution is also included. It is evident that there is greater consistency in \nmx determinations compared to $\Delta\nu$. 

The largest disagreements (beyond 2 $\sigma_{\rm{comb.}}$) typically occur at \nmx around the position of the clump ($20-30$ $\mu$Hz), highlighting an area of inconsistency between the different methodologies. Increased discrepancies are expected in this regime though, as the frequency spectra of core helium burning stars show more complex mode patterns and therefore parameter determinations are more dependent on the analysis techniques used. 


The \dnu distributions show a larger degree of scatter, as evidenced by their greater standard deviations. The scatter appears consistent across the range of values for each method, with the majority of values within twice the combined uncertainty of each other. 

Comparing the distributions to an N(0,1) distribution, it is clear that the standard deviations of the $\Delta(\nu_{\rm{max}})/\sigma$ distributions are all significantly lower than unity (see Fig. \ref{fig:seismo_comp}). This indicates that the independent analyses show strong agreement, but with a large correlation. The standard deviations of the $\Delta(\Delta\nu)/\sigma$ distributions are much closer to unity, showing good agreement, but a reduced correlation between methods. These results indicate that there is little disparity in the way \nmx is calculated for these data, but that the methods differ more in how they determine $\Delta\nu$. 

\subsection{Selection Function} \label{sel_fun}

The selection function used in this work was adapted from the K2 GAP proposals\footnote{All available at http://www.physics.usyd.edu.au/k2gap/.} for C3 and C6 (Sharma et al. \textit{in prep.}). The K2 GAP selection function was designed to be much simpler to implement than that for the \textit{Kepler} field \citep{2013MNRAS.433.1133F} and to ensure only red giants were observed. Its simplicity affords greater understanding of selection biases, and therefore trends, in the data. Cuts in colour and magnitude (JHK$_{\rm{s}}$ from the Two Micron All Sky Survey, \citealt{2006AJ....131.1163S}; V calculated from J and K$_{\rm{s}}$ as per C6 K2 GAP observing proposal\footnote{See http://www.physics.usyd.edu.au/k2gap/K2/C6 for proposal.}) are implemented within the K2 GAP for campaigns 3 and 6 as follows:

\begin{flushleft}
    \begin{equation}
    	\rm{C3}:
    	\begin{cases}
      		9.3 < \rm{V} < 14.5\\
      		\rm{J-K}_{\rm{s}} > 0.5\\
    	\end{cases}
    	\;\;\;\;\;
    	\rm{C6}: 
    	\begin{cases}
    		9 < \rm{V} < 15\\
    		\rm{J-K}_{\rm{s}} > 0.5\\
    	\end{cases}
    \end{equation}
\end{flushleft}

The V-band magnitude cuts differ between fields as the nominal cut for C3 was performed in the H-band ($7 < \rm{H} < 12$).
Further, to quantify whether the number of stars with detected oscillations follows the expectations, 
the asteroseismic detection probability of each star was calculated according to the method described in \cite{2019arXiv190110148S} and \cite{2011ApJ...732...54C}. In brief, the detection probability test takes an estimate of the seismic $\Delta\nu$ and $\nu_{\rm{max}}$, the granulation background and a theoretical Gaussian mode envelope for the star centred on the predicted $\nu_{\rm{max}}$. The signal-to-noise ratio of the estimated total mode power-to-background noise power within the envelope is used to determine the probability of an asteroseismic detection.

The seismic properties mentioned above are estimated using stellar radii and luminosities which were derived using a combination of astrometric and photospheric constraints. Luminosities were inferred from Ks magnitudes, bolometric correction from \citep{2014MNRAS.444..392C,2018ascl.soft05022C,2018MNRAS.475.5023C,2018MNRAS.479L.102C}, extinction calculated using the \code{mwdust} package \citep{2016ApJ...818..130B} with \cite{2015ApJ...810...25G} dust maps (extinctions are typically A$_{\rm{K}} < 0.05$ mag) and distances are based on Gaia DR2 data and the \cite{2016ApJ...832..137A} method. The input parallaxes were corrected for the zero-point offset based on their field location: $-15 \pm 4\mu$as for C3; $-2 \pm 2\mu$as for C6 (for calculation, see \citealt{2019arXiv190405676K}). The values of $T_{\rm{eff}}$ used were from the EPIC catalogue.

\begin{figure}
    \centering
    \includegraphics[width=\columnwidth]{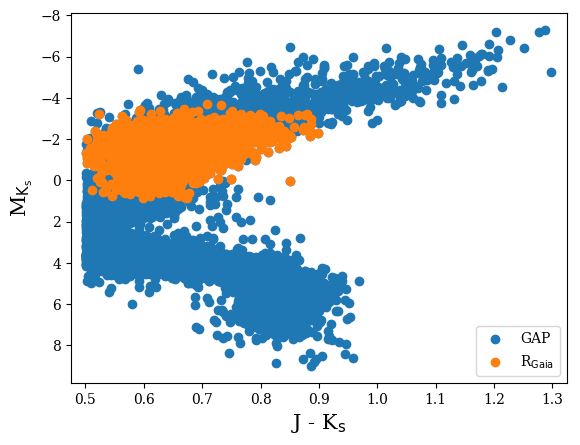}
    \caption{Colour-Magnitude Diagram for the C3 and C6 combined K2 GAP target lists. The blue markers indicate the original K2 GAP sample. Orange indicates the stars passing the detection probability test using radii derived from \textit{Gaia} parallaxes.}
    \label{fig:det_prob}
\end{figure}

The colour-magnitude diagram (CMD) in Fig. \ref{fig:det_prob} shows the distribution of all targets in the C3 and C6 fields registered in the K2 GAP target lists. Overlaid is the population of stars passing the detection probability test using radii calculated from the \textit{Gaia} parallaxes. All stars with calculated radii are grouped closely around the beginning of the red giant branch (RGB), the RGB bump (RGBb) and the red clump (RC). This is inline with the expectation of the selection function to remove all MS stars.

The number of stars from the K2 GAP predicted to pass the detection probability tests for C3 and C6 were 1073 and 1822 respectively. The number of stars with actual detections within the predicted sample were 762 for C3, and 1374 for C6. There is an $\sim 25\%$ reduction between the predicted and actual counts.  The reductions predominantly affect faint magnitude (H $> 10$, V $> 12$) stars. In C3, the initial star count with detections from the asteroseismic analysis is 885. Considering stars with both observational values for \nmx and $\Delta\nu$, 818 remain after the detection probability cut is made. Uncertainty in the estimate of the asteroseismic parameters for the K2 GAP sample is a possible source of the discrepancy, as are additional sources of instrumental noise not accounted for in the noise model.

A scaling relation was used to predict $\nu_{\rm{max}}$ for the K2 GAP sample for use in the probability test. Since stellar mass is not known, $\nu_{\rm{max}}$ was estimated from \textit{Gaia} radii,  $T_{\rm{eff}}$ from the EPIC catalogue, and
calibrated to the known K2 GAP $\nu_{\rm{max}}$ values from observations. The final form of the relation to estimate $\nu_{\rm{max}}$ was:

\begin{equation} \label{eq:nmx_beta}
    \nu_{\rm_{max}} \propto \left(RT_{\rm{eff}}^{-0.5}\right)^{-1.86}. 
\end{equation}

The values of $R$ and $T_{\rm{eff}}$ were varied within typical uncertainties, resulting only in a small ($< 10$ stars) difference between counts in each case. Hence we consider unlikely that inaccuracies in the predicted $\nu_{\rm{max}}$ are the main source of the discrepancy between the number of expected and returned detections.   

The noise model applied in the detection probability test follows that first proposed for \textit{Kepler} in \cite{2011ApJ...732...54C}, modified for the noise performance of K2 by \cite{2016PASP..128l4204L}.  The procedure predicts the expected global signal-to-noise level in the detected oscillations against background from intrinsic stellar noise (granulation) and shot/instrumental noise. This approach assumes that the population model is a good model of observations. Though the predictions work well for the majority of stars, in reality some show higher noise than expected (i.e. the tendency is for the distribution of real stars to show a high-noise tail), which could be a potential contributor to the lower actual numbers of detections returned by the different methods (see \citealt{Mosser_2019} for a review of factors affecting seismic performance).

After the detection probability test, a final cut was implemented based on observational detections. Stars with fewer than 2 independent detections were removed. Multiple detections were sought to improve the reliability of the measurement by cross-referencing. The values of $\nu_{\rm{max}}$ and $\Delta\nu$ were deemed consistent and the star accepted if the independent values lay within 2$\sigma$ of another. The final asteroseismic inputs were selected from the COR and BHM methods due to a greater yield of detections and less conservative uncertainties compared to A2Z. Though not explicitly used, the A2Z determinations are consistent with the final asteroseismic inputs.

Table \ref{tab:Data} states the final number of stars remaining from each survey after the selection cuts. The values associated with K2 are the final sample sizes after they have been cross matched with improved photometric metallicities and effective temperatures (see Section \ref{photom}).

\begin{table}
	\centering
	\caption{Number of stars in final samples for each K2 campaign field and supplementary surveys used in this work.}
	\label{tab:Data}
	\begin{tabular}{ccccc} 
		\hline
		Survey & C3 & C6\\
		\hline
		K2 & 483 & 929\\
        K2 SM & 377 & 646\\
        \hline
        RAVE & 85 & 83\\
        Gaia-ESO & 38 & -\\
        APOGEE & 101 & 25\\
		\hline
        K2 Spec. & 128 & 102\\
		\hline        
	\end{tabular}
\end{table}

\subsection{Photospheric Constraints} \label{phot_cons}

\subsubsection{Spectroscopy}

As with the asteroseismic data, multiple sources of spectroscopic data have been used to increase the yield of stars with such data. Once more, it is necessary to evaluate the consistency of these data to ensure a set of consistent results can be achieved independent of the spectroscopic survey used.

The spectroscopic data were collected to complement those from asteroseismology, improving upon the values of parameters such as $T_{\rm{eff}}$ and [Fe/H] that can be obtained from photometry. Data from the RAVE (C3/C6, \citealt{kunder_rave-gaia_2017}), APOGEE (C3/C6, DR16, \citealt{eisenstein_sdss-iii:_2011,majewski_apache_2017}; we make use of data to be released as a part of the Sloan Digitial Sky Survey IV, \citealt{2006AJ....131.2332G,2015AJ....150..148H,2015AJ....150..173N,2016AJ....151..144G,2017AJ....154...28B,2017AJ....154..198Z}) and Gaia-ESO (C3, Worley et al. \textit{in prep.}; \citealt{gilmore_gaia-eso_2012}) surveys have been used. After the selection cuts had been applied, these data were cross matched with the asteroseismic data. The final number of stars with spectroscopic information is low compared to the total sample (see Table \ref{tab:Data}), but still remains significant enough to draw sensible conclusions from the data and verify the wider trends observed with the larger photometric sample. Spectroscopic parameters from RAVE and Gaia-ESO were calibrated by adopting and iterating with asteroseismic log(g) values (\citealt{2017A&A...600A..66V}; Worley et al. \textit{in prep.}). Also, the log(g) values for the APOGEE sample are from asteroseismology, although the metallicities and $T_{\rm{eff}}$ are derived using the APOGEE spectroscopic log(g).

Though a thorough comparison between the surveys is beyond the scope of this paper, we bring attention to a source of potential bias within our results. We do not attempt to calibrate between surveys, but only demonstrate some differences between them.

The use of multiple sources of spectroscopic data is excellent for maximising the yield of targets, but comes with its own complexities. There is often little consistency between survey observations, with observations of different spectral domains and resolutions common \citep{2016AN....337..970V,2018arXiv181108041J}. In addition, each survey has a set of unique selection biases that need to be considered, which can manifest in systematic parameter trends (e.g. see \citealt{2018A&A...620A..76A,2018AJ....156..126J}). Cross-calibrating surveys thus proves difficult. Even where overlaps exist, it is not easy to directly compare the values \citep{2017ASInC..14...37J}.

\begin{figure*}
    \centering
    \includegraphics[width=\textwidth]{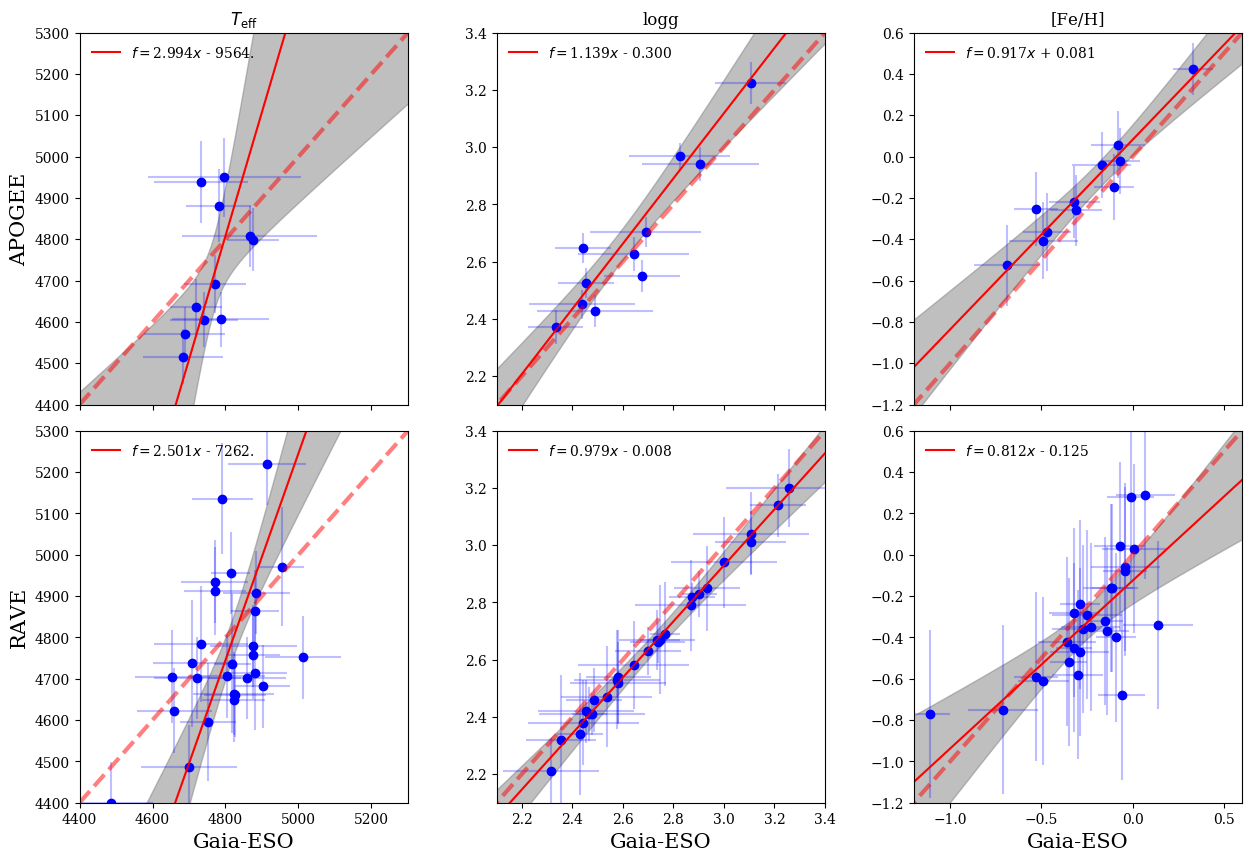}
    \caption{A comparison between the C3 spectroscopic data sources for $T_{\rm{eff}}$ (left panels), log(g) (middle) and [Fe/H] (right). Comparisons for APOGEE and Gaia-ESO (top row) and RAVE and Gaia-ESO (bottom row) are displayed. Blue points represent the data, with associated uncertainties. The red, dashed lines are the 1-to-1 relation to guide the eye. Red solid line is the best fit ($f$) to the data (equations given in legends) and the grey regions denote the confidence intervals of the fits.}
    \label{fig:spec_comps}
\end{figure*}

Comparisons and attempts to calibrate surveys to one another were made for this work in order to have a consistent spectroscopic sample. Figure \ref{fig:spec_comps} shows the comparison between the surveys used in C3, where an overlap of more than 10 stars was available (APOGEE to Gaia-ESO; RAVE to Gaia-ESO). The relations between $T_{\rm{eff}}$, log(g) and [Fe/H] are described by linear, orthogonal distance regressions (\code{odrpack, scipy}, \citealt{scipy2019}), with the resultant fits displayed on the relevant subplots. It is evident that though the uncertainties (APOGEE $\sigma_{\rm{Fe/H}}$ inflated by factor of 20) maintain consistency between values in each case, each gradient departs significantly from unity or is systematically offset. This is particularly strong in temperature and [Fe/H], revealing inconsistencies between the surveys. Consistent methodology to calibrate the log(g) values means reduced scatter, but a departure from unity is still observed.

Considering the scatter between surveys, prior to the modelling process the uncertainties on [Fe/H] and $T_{\rm{eff}}$ were increased to account for the observed differences. The variance between surveys was of order 50 K in $T_{\rm{eff}}$ and 0.1 dex in [Fe/H] respectively. These values were adopted as the systematic uncertainty between surveys and were added in quadrature to the initial survey values. The resultant values are conservative uncertainties for the spectroscopic parameters.



\subsubsection{Chemical Composition}

Chemical space is often a key area used in the literature to distinguish between stars belonging to a thin or thick disc population. Typically, the thin disc is expected to be [Fe/H] rich and solar-[$\alpha$/Fe]; the thick disc [Fe/H] poor and [$\alpha$/Fe] enriched (e.g. see \citealt{1998A&A...338..161F,2005A&A...433..185B,2007ApJ...663L..13B,2008MNRAS.391...95R,2011ApJ...737....9R,2015A&A...582A.122K}). This trend is a consequence of the expected epochs of formation of these structures. The thick disc is considered to be older (10-12 Gyr) and have therefore formed rapidly in conditions with less metal enrichment and greater $\alpha$-enhancement from core-collapse supernovae. The thin disc is thought to have started forming later (7-9 Gyr ago) and, as a consequence, is more metal-rich due to enrichment of the interstellar medium by type-Ia supernovae 
(see e.g. \citealt{2001ASSL..253.....M} and references therein).

\begin{figure}
	\centering
    \includegraphics[width=\columnwidth]{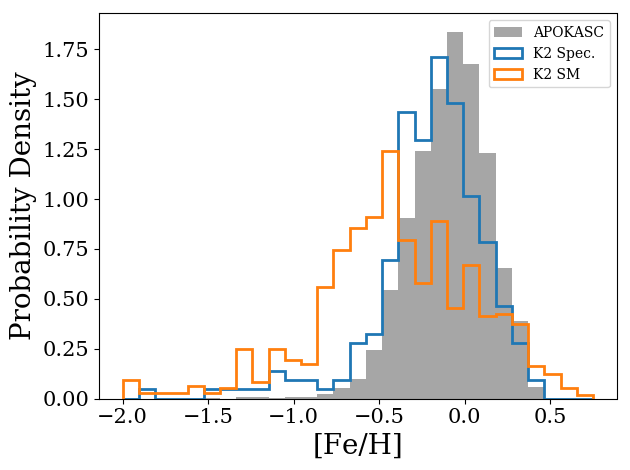}
    \caption{Normalised [Fe/H] distributions for the APOKASC (grey), K2 stars with spectroscopic values (blue) and SkyMapper values (orange).}
    \label{fig:FeH_dist}
\end{figure}

The metallicity distributions observed with \textit{Kepler} and the K2 fields studied here further demonstrate the suitability of the K2 fields for this study and are shown in Figure \ref{fig:FeH_dist}. The \textit{Kepler} distribution peaks at [Fe/H] of -0.1 dex with a standard deviation of $\pm0.5$ dex. This distribution is highly indicative of a thin disc dominated population, with only a small tail in the metal poor regime. A peak for the thick disc would be expected at around -0.5 dex (see \citealt{2013A&A...558A...9M,2014A&A...572A..92M} and references therein). The [Fe/H] values have typical uncertainties of 0.21 dex.

The K2 Spec. distribution follows closely that of the APOKASC sample, though it peaks at a lower metallicity (-0.25 dex). The whole sample is clearly shifted towards lower metallicity and has an extended metal-poor tail. Given the greater vertical extension of K2 C3 and C6, the sample is likely to have a dominant contribution from the thick disc, explaining the shift compared to APOKASC. The K2 SM sample alludes to a much greater metal poor tail than K2 Spec. shows. This is potentially true, but the photometric distribution has greater scatter compared to the RAVE and APOGEE survey metallicities (see \citealt{2019MNRAS.482.2770C}). This is demonstrated by the extension to unlikely metallicities of $> 0.5$ dex for K2 SM.



\begin{figure}
	\centering
    \includegraphics[width=\columnwidth]{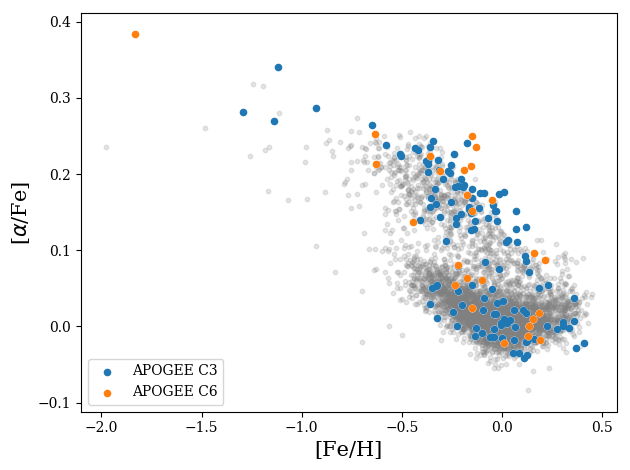}
    \caption{[Fe/H] vs. [$\alpha$/Fe] distribution for the APOGEE survey in the K2 C3/C6 samples. The APOKASC distribution is shown in grey. The cross shows the typical uncertainties in [Fe/H] and [$\alpha$/Fe] in the sample used.}
    \label{fig:alpha_fe}
\end{figure}


The [Fe/H] vs. [$\alpha$/Fe] distribution for the K2 APOGEE and APOKASC samples is shown in Fig. \ref{fig:alpha_fe}. The same spectroscopic survey has been used here for a direct comparison to negate the effect of any biases due to the survey selection function. The typically expected low- and high-$\alpha$ sequences associated with the evolution of the thin and thick discs are evident for these populations. Classifying the $\alpha$-rich population as in Table \ref{tab:cross_refs} ([$\alpha$/Fe] $> 0.1$), $\sim 60\%$ of the K2 C3/C6 sample consists of $\alpha$-rich stars compared to only $\sim 15\%$ of the APOKASC sample. It is thus necessary to include comparisons of the K2 samples to the $\alpha$-rich component of APOKASC and not only the full sample. Though the proportion of stars in this regime is smaller than in the K2 populations, the sample size is still significant enough  (748 stars) for comparisons and conclusions to be drawn.

\subsubsection{Photometry} \label{photom}

Spectroscopic temperatures and metallicities were not available for all of the stars in the initial sample. To supplement this information, photometric values of $T_{\rm{eff}}$ and [Fe/H] determined from observations by the SkyMapper survey \citep{2007PASA...24....1K,2018PASA...35...10W} have been used. SkyMapper is designed to take \textit{uvgriz} photometry, from which stellar parameters can be derived \citep{2019MNRAS.482.2770C}. The reported survey parameters have been calibrated using GALAH \citep{2018MNRAS.478.4513B} spectroscopic metallicities and $T_{\rm{eff}}$ from the InfraRed Flux Method, and validated against APOGEE DR14 \citep{2018ApJS..235...42A} and RAVE DR5 to ensure reliable parameter determinations. 

The SkyMapper survey has covered most of the southern sky, subsequently observing stars across both C3 and C6. This provides a coherent photometry source for the fields and parameter determinations. Parameters were not available for all of the stars in the two campaigns, but the total number of stars remains significant for understanding population trends (377 - C3; 646 - C6). As they number fewer, the stars with spectroscopic values are considered a sub-sample of the total photometric population in this work. The spectroscopic values should provide tighter constraints on the final parameter distributions and better information regarding the chemistry of the stars than the photometry considering the higher-resolution observations. These are therefore viewed as a benchmark to which the larger photometric sample can be compared and underlying trends identified/ratified.

\subsection{Synthetic Data}

In addition to using \textit{Kepler} APOKASC data to provide a comparative sample for the K2 data, a synthetic population was also generated using the TRILEGAL (a TRIdimensional modeL of thE GALaxy, \citealt{2012ASSP...26..165G}) population synthesis code. The code allows one to generate a synthetic population of a specific region of the Galaxy, with user defined Galactic components, initial mass function and star formation rates. We computed two synthetic populations based on the central coordinates of the C3 and C6 fields and the field of view of the \textit{Kepler} telescope. A uniform star formation rate from 9 Gyr to present was chosen for the thin disc and a 1 Gyr burst at 11 Gyr for the thick disc.

\section{Method} \label{method}

We use grid based stellar modelling to extract the fundamental parameters of the stellar ensembles. A grid of models generated using MESA (Modules for Experiments in Stellar Astrophysics; \citealt{2015ApJS..220...15P}) in conjunction with the Bayesian inference tool \textsc{PARAM} \citep{2006A&A...458..609D,2014MNRAS.445.2758R,2017MNRAS.467.1433R} were used. 

From the measured observational constraints  ($T_{\rm{eff}}$, [Fe/H]),  $\nu_{\rm{max}}$, and $\Delta\nu$) \textsc{PARAM} computes probability distribution functions for the stellar parameters (e.g. radius, mass, age).  The code uses a flat prior age and an initial mass function from \citet{Chabrier2001}. An upper age prior can also be set, which we set as 20 Gyr. Though greater than the accepted age of the Universe,current uncertainties on stellar ages are typically greater than $30\%$. Ages up to 20 Gyr are consistent with being drawn from a normal distribution centred on the Hubble age with $\sigma = 30\%$.

Parameter values are determined statistically from the output probability density functions (PDF) produced by \textsc{PARAM} \citep{2017MNRAS.467.1433R}. A choice of using the median or mode statistic is available to the user. The 68th and 95th percentiles are returned for all parameters in each case. 


We adopt the modal values as the preferred choice of final parameter. The modal value is most representative of the distribution peak, particularly when approaching the limits of the underlying grid boundaries and priors. 
Post process, stars caught on the prior boundaries of the grid in age are removed from the sample. These stars are forced to specific ages, potentially distorting the final parameters returned. An upper limit to the prior of 20 Gyr was used. This is reflective of the age distribution observed in Fig \ref{fig:Age_KDE}, whereby stars with high uncertainty are attributed ages beyond the expected age of the Universe. The final sample sizes of C3 and C6 are 377 and 646 respectively (10$\%$ and 6$\%$ reduction compared to values in Table \ref{tab:Data}).

\section{Radii} \label{radii}

\begin{figure}
	\centering
    \includegraphics[width=\columnwidth]{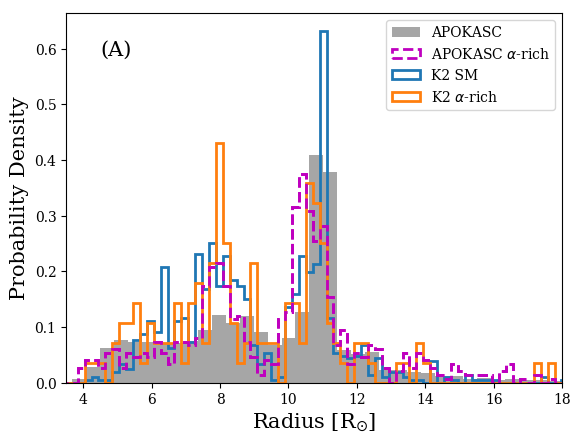}
    \includegraphics[width=\columnwidth]{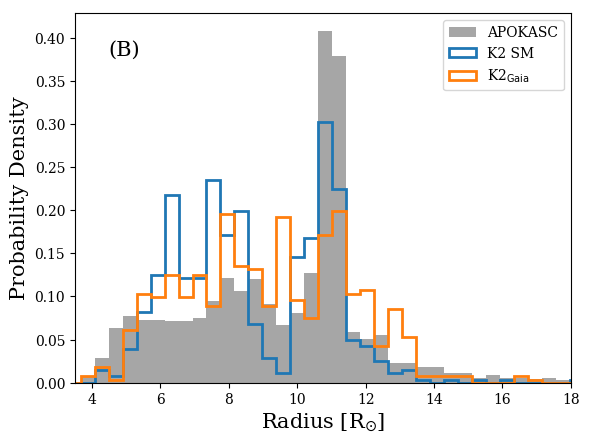}
    \includegraphics[width=\columnwidth]{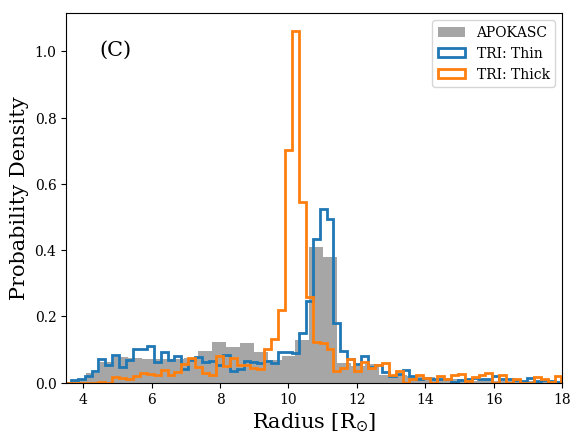}
    \caption{\textit{(A):} Radius distributions from \textsc{PARAM} for the APOKASC (grey), APOKASC $\alpha$-rich (magenta, dashed), K2 SM (blue) and the K2 $\alpha$-rich (orange) samples. \textit{(B):} Comparison of \textit{Gaia} (magenta) and seismic (blue, \textsc{PARAM}) radii distributions for the same stars in K2 SM. Only stars with $\sigma_{\varpi}/\varpi < 10\%$ are shown. \textit{(C):} TRILEGAL simulation of the K2 C3 field. The thin (blue) and thick (orange) disc populations within the simulation are shown. All distributions are normalised.}
    \label{fig:Radii}
\end{figure}

An examination of the distribution of radii within the K2 sample provides a good indicator if a typical population of red giant stars is observed. Though the distribution will vary between observed populations, key features such as the red clump should be obvious from a pronounced peak at $\sim10-11$ R$_{\odot}$. Figure \ref{fig:Radii} displays this characteristic, indicating the clump sample within the data. The figure shows a comparison of \textsc{PARAM} (i.e. asteroseismic) radii for K2 SM and APOKASC (panel A) and \textsc{PARAM} to \textit{Gaia} K2 SM (panel B; $\sigma_{\varpi}/\varpi < 10\%$ cut applied) radii. Comparing K2 to APOKASC provides context for the results, but comparing the radii derived using \textsc{PARAM} and those computed using \textit{Gaia} parallaxes provides insight into which values are most appropriate to use in future analyses. The \textit{Gaia} distribution at the red clump peaks at a lower radius than that from \textsc{PARAM} for the same stars. As discussed in section \ref{sel_fun}, this shows evidence of an under-estimation of the stellar radii compared to asteroseismology.


\begin{figure*}
	\centering
	\begin{tabular}{cc}
	\includegraphics[width=\columnwidth]{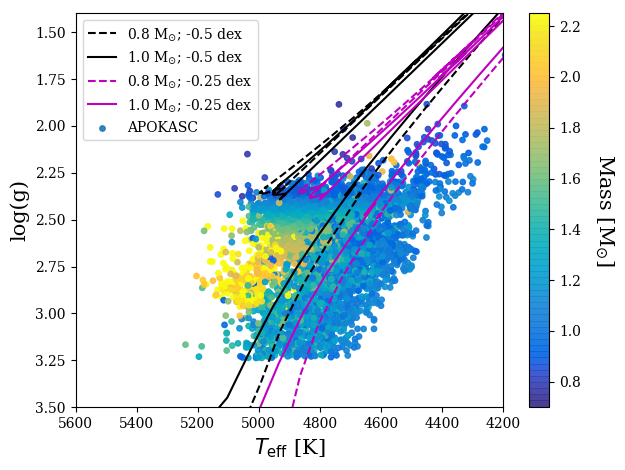} &
	\includegraphics[width=\columnwidth]{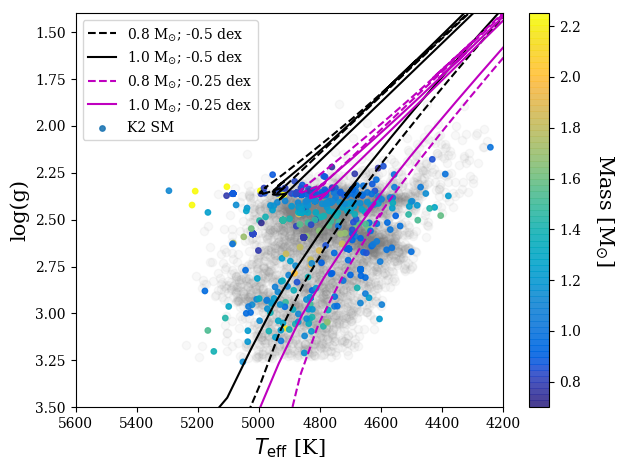} \\
	\end{tabular}
	\includegraphics[width=\columnwidth]{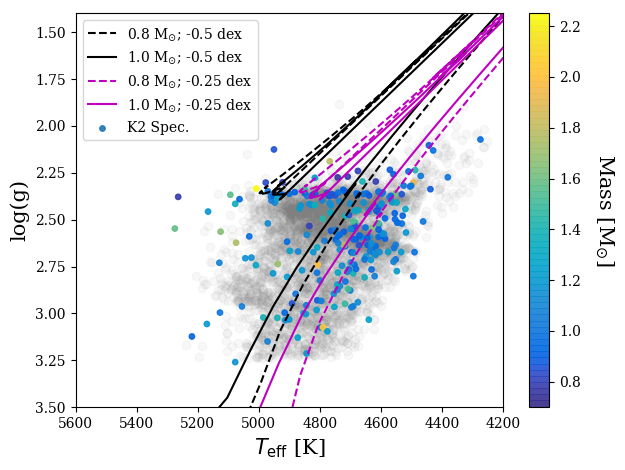}
    \caption{Kiel diagram with mass colour bar for the APOKASC (top left), photometric (top right) and spectroscopic (bottom) K2 samples. Tracks of different mass (0.8 M$_{\odot}$ - dashed; 1.0 $_{\odot}$ - solid) and metallicity ($-0.5$ dex - black; $-0.25$ dex - magenta) are overlaid as a guide. The $\alpha$-rich APOKASC population is included in grey. Grey lines denote the location of the RGB-bump (RGBb).}
    \label{fig:ASK}
\end{figure*}

An unexpected secondary peak at $\sim7-8$ R$_{\odot}$ is present in both the spectroscopic and photometric K2 data. The secondary peak is also a feature of a sample analysed in Miglio et al. (in prep.). The sample is a population of $\alpha$-rich ([$\alpha$/Fe] $> 0.1$) stars from the APOKASC catalogue. This $\alpha$-rich sample shows comparable features to those observed with the K2 stars, indicating that the feature is typically common to an older population, a trait synonymous with $\alpha$-enhanced stars. A detailed examination of a Kiel diagram (Fig. \ref{fig:ASK}) of the populations shows an overdensity of stars located at the RGB-bump, which is synonymous with the over density in radius observed.

\begin{figure}
    \centering
    \includegraphics[width=\columnwidth,keepaspectratio]{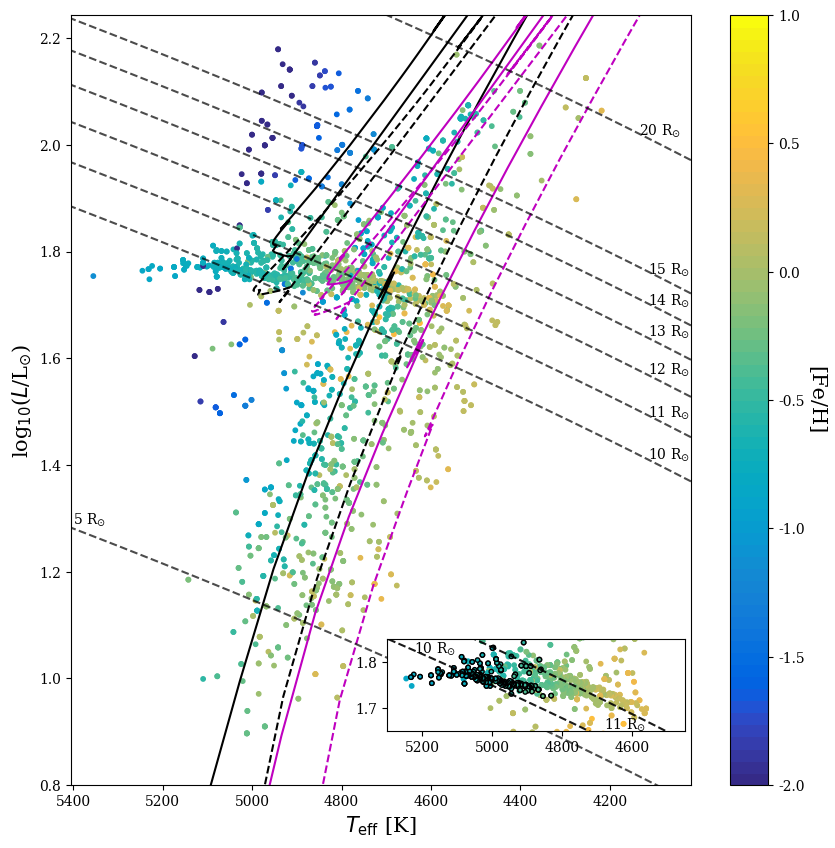}
    \caption{HRD of a K2 C3 TRILEGAL simulation. An [Fe/H] colour bar has been used, with lines of constant radius included (grey dashed line; values marked on plot). Tracks of different mass (0.8 M$_{\odot}$ - dashed; 1.0 $_{\odot}$ -solid) and metallicity (-0.5 dex - black; -0.25 dex - magenta) are overlaid as a guide. \textit{Inset:} An enlargement of the red clump population, with stars classified as thick disc in the simulation denoted by black circles. Only red clump stars are shown here, using the classifier in TRILEGAL.}
    \label{fig:rad_tri}
\end{figure}

Interestingly, there is a slight difference in the peak of the clump distributions of the full and $\alpha$-rich samples for both APOKASC and K2. The clump distribution for the $\alpha$-rich APOKASC sample peaks 0.5 R$_{\odot}$ lower than the full population. More metal poor stars were found to have lower radii. This difference is less significant ($\sim$0.1 R$_{\odot}$) in K2. Simulations from TRILEGAL predict that this is also related to the expected Galactic component that a star is a member of. The clump of the thick disc peaks at the observed $\alpha$-rich radii and the thin disc at the full APOKASC radii as one may expect (see Fig. \ref{fig:Radii}, panel C). The trend is a function of $T_{\rm{eff}}$, driven by mass and metallicity. Evidence of this can be seen in Fig. \ref{fig:rad_tri}. Here, a Hertzsprung-Russell diagram, coloured by metallicity, of a K2 C3 TRILEGAL simulation is shown. It is evident that there is a division in the red clump population, whereby more metal poor stars are situated at hotter temperatures, hence, a lower radii than more metal rich stars, given that the luminosity is very similar. This is highlighted further by the inset of the Fig. \ref{fig:rad_tri}, which shows a zoom in of the red clump population. The overlaid evolutionary tracks (as in Fig. \ref{fig:ASK}) show this as a function of mass too. The hot clump sample ($>$ 4900 K) lies close to the low mass and metallicity track (0.8 M$_{\odot}$; -0.5 dex), whereas the cooler clump sample ($< 4900$ K) is positioned closest to the high mass and metallicity track (1 M$_{\odot}$; -0.25 dex). 

A separation in the populations in radius is apparent (though some contamination from first ascent RGB stars is present), further confirming the trends observed in Fig. \ref{fig:Radii} and Fig. \ref{fig:rad_tri}. This is not conclusive evidence that the positioning of the peak clump radius can be used as a tracer for Galactic components/evolution, but opens up possibilities to explore this further (see \citealt{2016ARA&A..54...95G} and references therein).

\section{Masses} \label{mass}

\begin{figure*}
	\centering
	\begin{tabular}{cc}
	\includegraphics[width=0.48\textwidth]{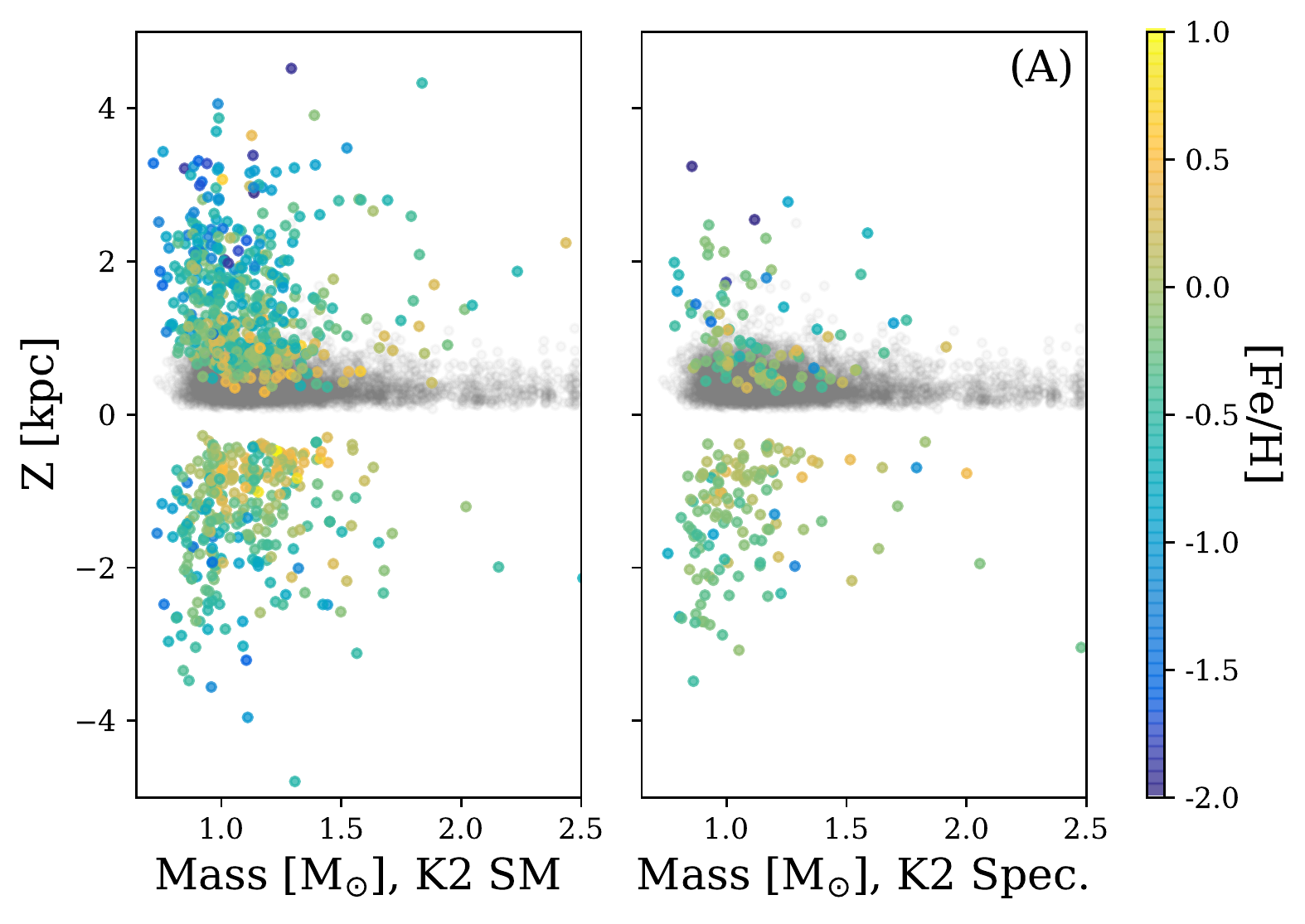} & 
	\includegraphics[width=0.48\textwidth]{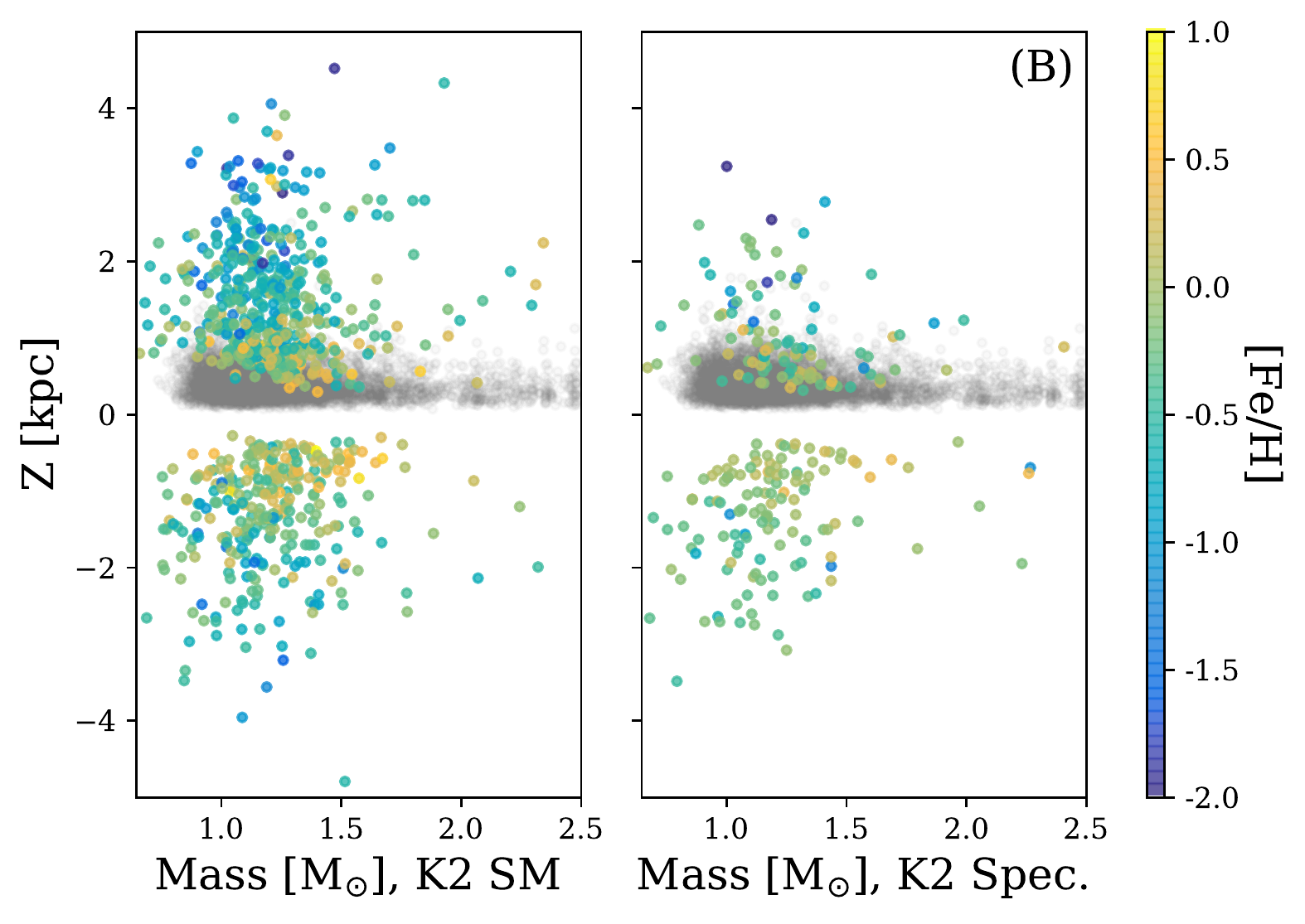}\\
	\includegraphics[width=0.48\textwidth]{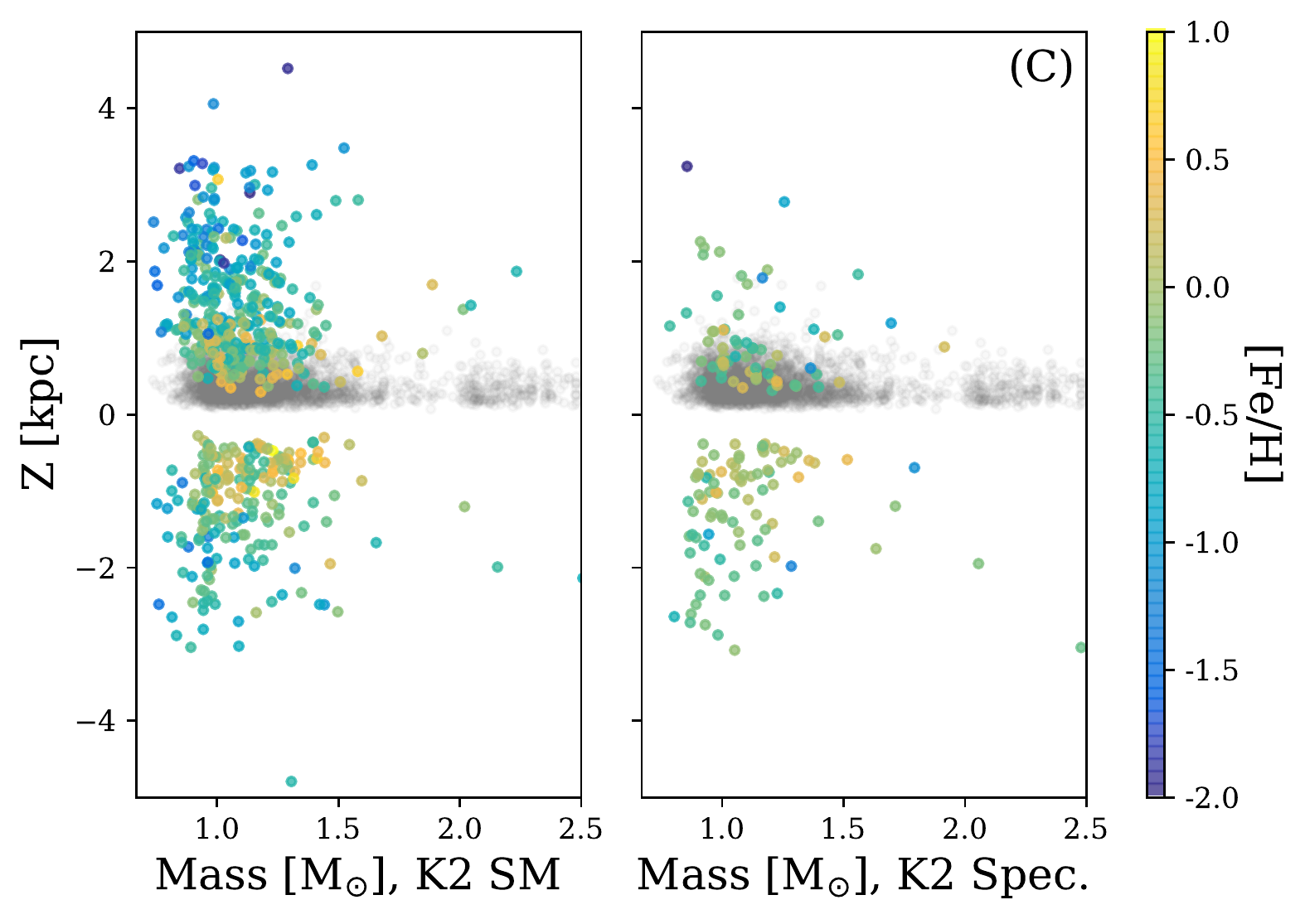} &
	\includegraphics[width=0.48\textwidth]{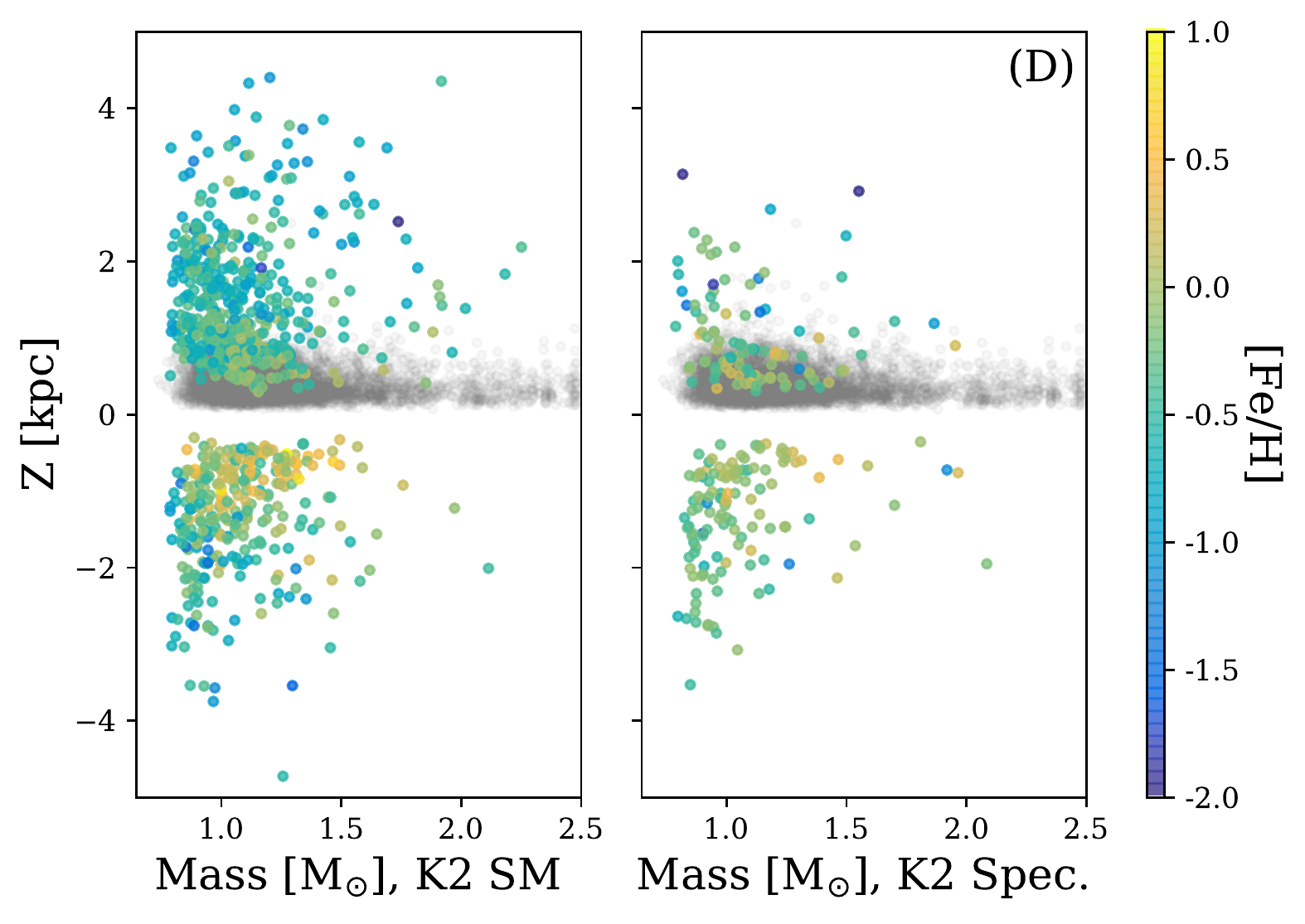}\\
    \end{tabular}
	\caption{Mass against vertical height above and below the Galactic plane ($Z$) for the K2 SM (left) and K2 Spec. (right) samples in each panel. An [Fe/H] colour bar is shown. The metallicity scale is the same for each subplot. The APOKASC sample is shown in grey. \textit{(A):} Original K2 SM and K2 Spec. samples. \textit{(B):} Masses calculated from scaling relations. \textit{(C):} Red clump population has been removed. \textit{(D):} Populations re-analysed with a grid including diffusion.}
    \label{fig:MZF}
\end{figure*}

Given the tight age-mass relation expected for red giant stars \citep{Kippenhahn}, stellar masses inferred by asteroseismology provide an excellent proxy for age \citep[e.g., see ][]{Miglio2012, 2016AN....337..774D}. Understanding the mass distribution of a population therefore allows early inferences about the expected age distribution to be made. Panel A of Fig. \ref{fig:MZF} shows the mass distribution of the K2 SM, K2 Spec. and APOKASC samples as a function of vertical height, $Z$, from the Galactic plane. It can be seen that all samples show a trend of increasing vertical extent with decreasing mass. There is evidence of a metallicity gradient, with decreasing metallicity observed as one moves out of the plane and towards lower masses. These trends are comparative with those expected of a thin/thick disc structure, in particular with the low masses extending to greater vertical extent being reflective of the expectation of observing older stars further from the plane (see section \ref{ages}; \citealt{2013MNRAS.429..423M, 2016MNRAS.455..987C}). 

\begin{figure}
    \centering
    \includegraphics[width=\columnwidth]{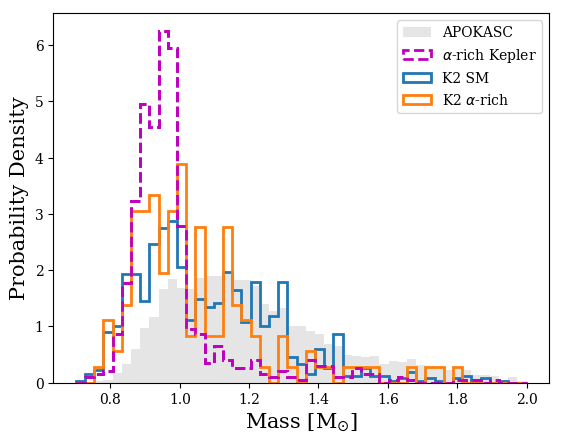}
    \caption{Normalised mass distributions from \textsc{PARAM} for the APOKASC (grey), APOKASC $\alpha$-rich (magenta, dashed), K2 SM (blue) and the K2 $\alpha$-rich (orange) samples.}
    \label{fig:mass_distr}
\end{figure}

The remaining panels of Fig. \ref{fig:MZF} show the resultant populations after additional cuts/re-analyses of the K2 SM and K2 Spec. data were made (same alterations applied to the background APOKASC samples). The effects due to using masses from scaling relations \citep{1995A&A...293...87K} (B), removing the red clump by radius (C) and using a grid including microscopic diffusion (D) were explored to test for any property dependencies within the populations. Except additional scatter at low masses when the scaling relations are used (a likely over-estimation of the masses; see e.g. \citealt{2017MNRAS.467.1433R} and references therein) 
and varying sample sizes, the initial trends seen in panel A are invariant to the changes implemented. The high mass scatter and shape of the mass/$Z$ relations remain consistent throughout, as does the perceived metallicity gradient. The robustness of these trends gives confidence to the derived stellar properties in the K2 SM and K2 Spec. samples being a true reflection of the population and its features.

Figure \ref{fig:mass_distr} shows the mass distributions of both the full and $\alpha$-rich APOKASC and K2 populations. The distributions show that the K2 SM sample contains a larger proportion of low mass stars than in APOKASC, suggesting that the population of these K2 fields is potentially older than that of APOKASC and is discussed further in section \ref{ages}. It is also shown that the number of $\alpha$-rich stars in these populations is enhanced for lower masses, suggesting the older a red giant star is, the more $\alpha$-enhanced it is likely to be.

\section{Ages} \label{ages}

Many arguments surrounding the definition of the thin and thick discs, in particular their formation, centre largely on the age distribution of the populations and indications of enhanced star formation. Conclusively proving a distinction in age between the populations defined geometrically and/or chemically is difficult due to typical age uncertainties of $>40\%$. The samples used in this work have median uncertainties of $< 35\%$ (X$_{\rm{HQ}}$ samples), allowing general trends to be extracted. 

\begin{figure*}
	\centering
    \includegraphics[width=\textwidth]{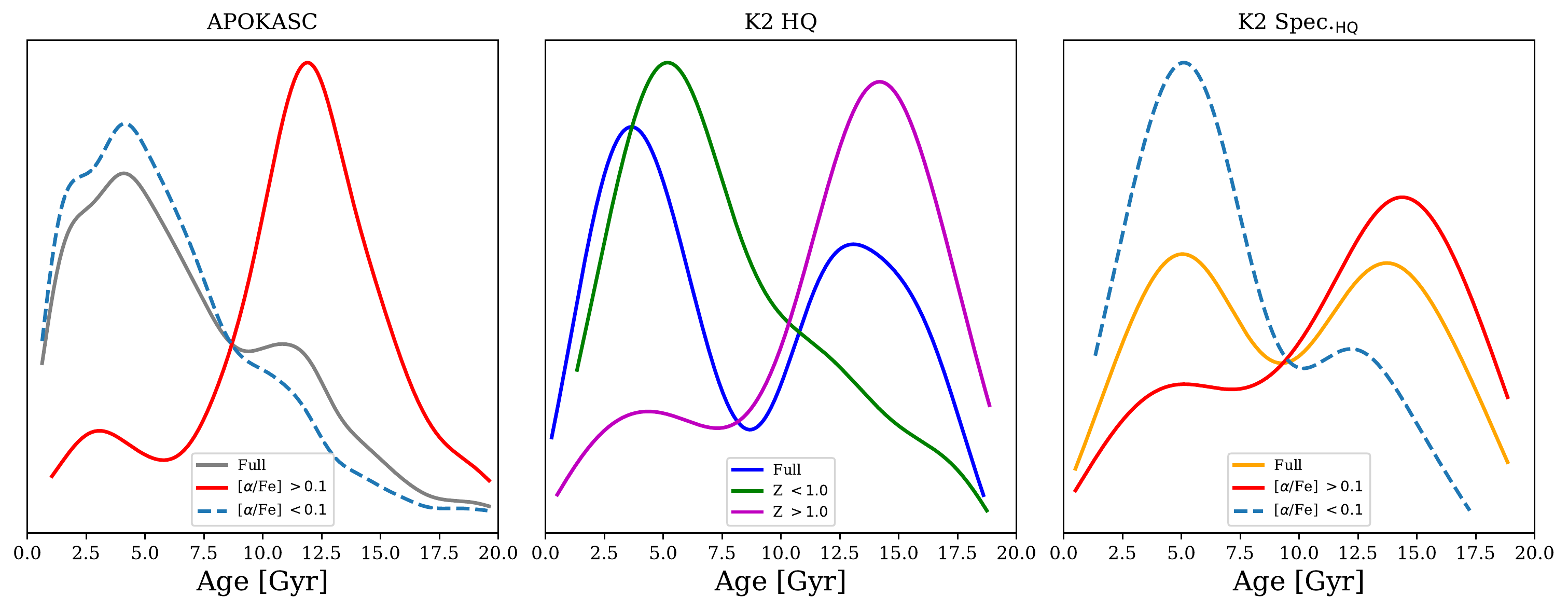}
    \caption{Normalised age distributions for the APOKASC and K2 populations. \textit{Left:} Nominal (grey), $\alpha$-rich (red) and $\alpha$-poor ([$\alpha$/Fe]$< 0.1$; blue dashed) APOKASC sample age distributions. \textit{Middle:} K2$_{\rm{HQ}}$ complete (blue), $|Z| < 1.0$ kpc (green) and $|Z| > 1.0$ kpc (purple) distributions. \textit{Right:} K2 Spec.$_{\rm{HQ}}$ (orange) and K2-APOGEE $\alpha$-rich (red) and $\alpha$-poor ([$\alpha$/Fe]$< 0.1$; blue dashed) distributions.}
    \label{fig:AgeKDE}
\end{figure*}

Figure \ref{fig:AgeKDE} shows a comparison of the age distributions\footnote{Kernel Density Estimates were generated using the \code{python} module \code{PyQt-Fit}, \code{1DKDE}. Default smoothing applied.} of the APOKASC (Miglio et al., \textit{in prep.}), K2$_{\rm{HQ}}$ and K2 Spec.$_{\rm{HQ}}$ samples for stars with age uncertainties $< 35\%$. Each set of stars was analysed with an extended age prior of 20 Gyr, hence the un-physically extended age ranges. As previously stated, the magnitude of the uncertainties are such that these values remain consistent with the age of the Universe within 1-2 $\sigma$. It is the features of the distributions that are of most interest though. The APOKASC population follows closely the age distribution for giants shown by \cite{2016MNRAS.455..987C} with a large peak at 5 Gyr and a smaller peak at 11 Gyr. In addition, the $\alpha$-rich population maps closely the distribution presented in \cite{2018MNRAS.475.5487S}. The distribution peaks broadly at $\sim 12$ Gyr, with a small overdensity at 3 Gyr due to a population of young $\alpha$-rich stars (e.g. see \citealt{2011MNRAS.414.2893F,2015A&A...576L..12C,2015MNRAS.451.2230M,2016AN....337..917J,2016A&A...595A..60J}). The dominance of the 12 Gyr peak in the $\alpha$-rich population indicates that this influences the appearance of the secondary peak in the nominal APOKASC sample. A reduced prominence at 12 Gyr for the $\alpha$-poor ([$\alpha$/Fe]$< 0.1$) APOKASC sample confirms this, reaffirming the expectations that older stellar populations have enhanced $\alpha$-element abundances. The consistency of our results with these studies gives confidence to make clear comparisons with results from K2. 

Considering the APOKASC and K2 populations extend vertically beyond 1 kpc, it is expected that some mixing will occur between disc populations as the thin disc transitions up into the thick disc. Hence, both samples should contain a prominence related to each component. The broadness of the peaks is not of concern here as it is known that the K2 data is not as high quality as the \textit{Kepler} data and therefore greater uncertainties are expected. Figure \ref{fig:Age_KDE} shows the impact of different age uncertainties on the shape of a simulated distribution. Large uncertainties mask the original features, emphasising the importance of obtaining high-precision age determinations.

The K2$_{\rm{HQ}}$ photometric and K2 Spec.$_{\rm{HQ}}$ distributions are shown in the centre and right hand panels of Fig. \ref{fig:AgeKDE}. The K2$_{\rm{HQ}}$ distribution peaks predominately at 5 Gyr in concordance with APOKASC. The distribution then passes through a minimum at $\sim 9$ Gyr before gradually increasing again towards older ages. This differs slightly to the spectroscopic sample, which shows an earlier minimum at 8 Gyr (as with APOKASC) and a clearly defined secondary peak at 14 Gyr. Given the current uncertainties on age and the fact that the two samples differ both in magnitude range and in photospheric parameters (e.g. for the stars in common the spectroscopic temperatures are on average $\sim 50$ K lower than SkyMapper's), these small apparent age differences are likely due to be dominated by target selection biases and systematic effects. 
K2 Spec.$_{\rm{HQ}}$ also shows a more even weighting between the young and old peaks, suggesting a split population. As previously discussed, the K2 age uncertainties are larger compared to APOKASC, which contributes to the overall washing out and extension of features in the distributions. The greater defined features of the spectroscopic sample compared to the photometric is symptomatic of the quality of the input parameters used in the analysis, with the spectroscopic surveys providing improved input parameters and, hence, final uncertainties. Similarly to with APOKASC, the $\alpha$-rich and $\alpha$-poor spectroscopic components from APOGEE were plotted. It is clearly shown that the young and old peaks are dominated by the $\alpha$-poor and $\alpha$-rich sub-samples respectively. This further confirms the expected chemical dichotomy of stars from different generations.

The higher age peak, and larger proportion of stars at older ages when using spectroscopic inputs, indicates that the K2 fields contain an older population than the \textit{Kepler} sample; but is this just a radial selection effect or due to sampling a greater vertical extent? Considering the majority of the APOKASC sample extends out to 1 kpc, the population trends for the K2$_{\rm{HQ}}$ stars above and below $\pm$1 kpc are also shown on the centre panel of Fig. \ref{fig:AgeKDE}. The sample below 1 kpc follows closely that of the APOKASC sample, with a very clearly defined young population at 5 Gyr, but has few stars beyond 10 Gyr, with the distribution dropping off significantly and flattening beyond 9 Gyr. In contrast, the population beyond 1 kpc shows a minimal peak at $\sim$5 Gyr and dominant old population peak at 15 Gyr. The difference in the shape of the age distributions above and below 1 kpc is stark enough to show that the stellar population changes with increasing $Z$. 

\begin{figure}
	\centering
    \includegraphics[width=\columnwidth]{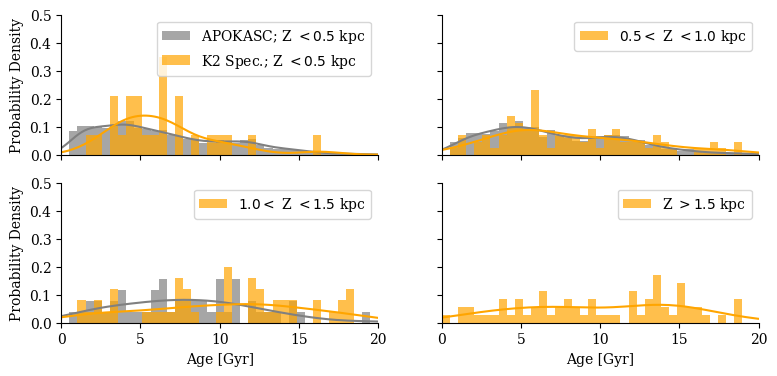}
    \caption{K2 Spec.$_{\rm{HQ}}$ sample age distribution as a function of $Z$ (0.5 kpc bins). A comparison to the APOKASC sample is performed up to 1.5 kpc as beyond this the numbers are insufficient for a meaningful comparison. Samples of stars with $\sigma_{\rm{age}} < 35\%$ shown.}
    \label{fig:AZ}
\end{figure}

For further confirmation that the age distribution changes primarily with $Z$ and not with Galactic radius, Fig. \ref{fig:AZ} shows the age distribution of the K2$_{\rm{HQ}}$ field and APOKASC samples in vertical bins of 0.5 kpc. The C3 field was chosen as this field samples significantly beyond 1.5 kpc after all cuts are applied, whereas C6 contains too few stars beyond 1.5 kpc for sensible conclusions. For the same reasoning, the APOKASC sample is only shown out to 1.5 kpc. The first two bins of Fig. \ref{fig:AZ} (0 - 1 kpc) show that the APOKASC and K2 populations follow each other closely. A two sample Kolmogorov-Smirnoff test confirms this consistency with $p$-values greater than 0.05 for each bin ($p_{Z<0.5} = 0.28$, $p_{0.5<Z<1.0} = 0.09$), rejecting the hypothesis that the APOKASC and K2 C3 population age distributions are significantly different in these ranges. This is a good indication that, up to 1 kpc, the age distribution of similarly selected stars is expected to be the same at these different Galactic radii. Consequently, any further inferences can be concluded to be due to the vertical rather than radial properties of the fields. 

\begin{figure}
	\centering
    \includegraphics[width=\columnwidth]{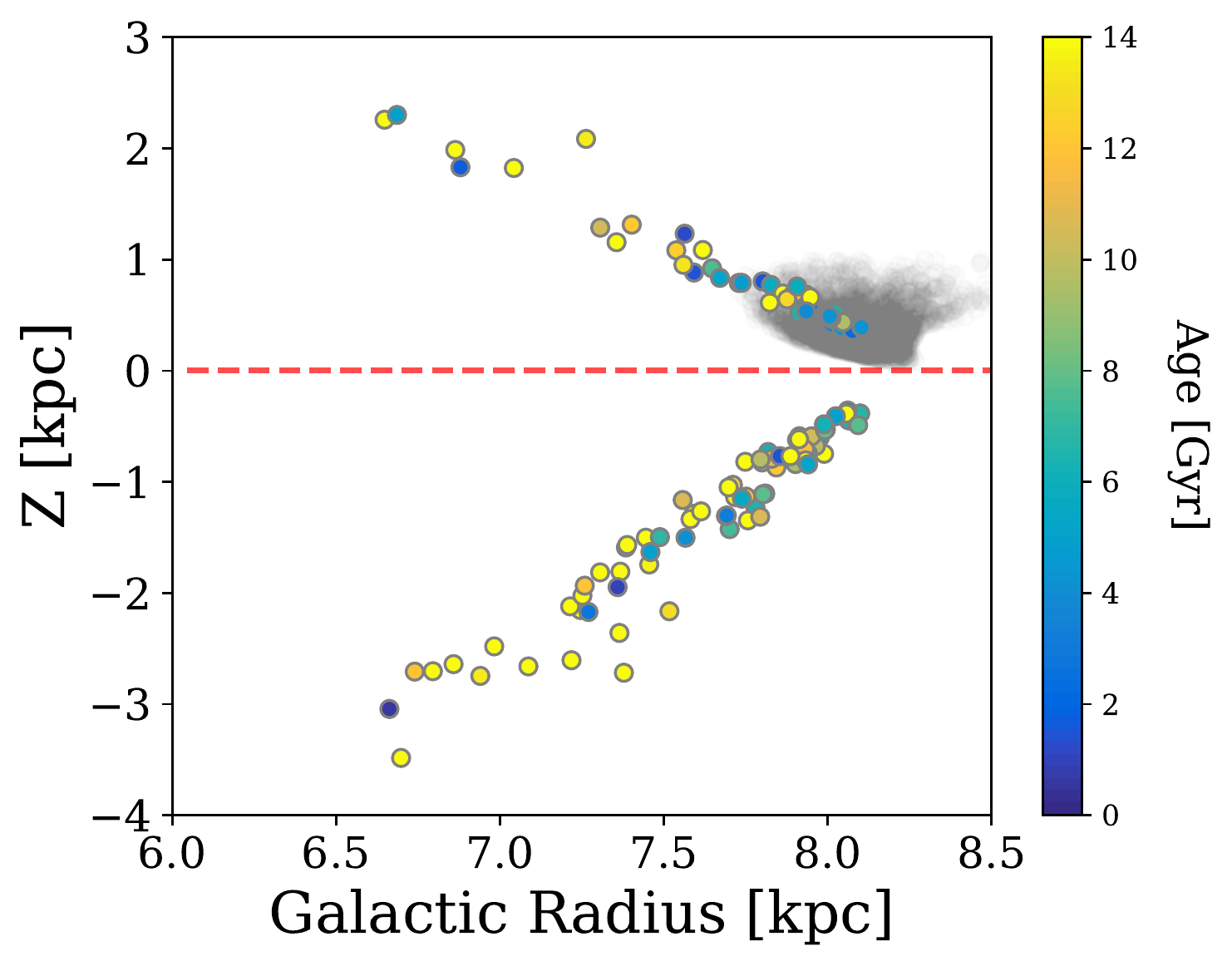}
    \caption{As Fig. \ref{fig:RZ_dist}, but showing the age distribution of the K2 Spec.$_{\rm{HQ}}$ sample.}
    \label{fig:RZA}
\end{figure}

Reflecting the trend observed in Fig. \ref{fig:AgeKDE}, the bins beyond 1 kpc show an increasingly divergent population, with consistent young and old populations. This was not expected due to the typical belief that the thick disc is composed of older stars. An explanation of this is the presence of the aforementioned young ($< 7$ Gyr), $\alpha$-rich stars. Considering the K2 Spec., $\sigma_{\rm{age}} < 35\%$ sample, $\sim 8\%$ of the stars (accounting for increased intrinsic uncertainties on age compared to \textit{Kepler}) fall into this category, compared to $\sim2\%$ in APOKASC (same $\sigma_{\rm{age}}$ cut). Of these stars, 20$\%$ can be found beyond 1 kpc from the Galactic mid-plane. This indicates an expectation to see non-insignificant numbers of these stars at high $|Z|$. Fig. \ref{fig:RZA} illustrates this trend. The figure is a replication of Fig. \ref{fig:RZ_dist}, but only for the K2 Spec.$_{\rm{HQ}}$ sample and is coloured by age. It is readily apparent that the older population stars ($> 10$ Gyr) dominate at high-$|Z|$ ($> |1.5|$ kpc), but contamination by young stars is clearly visible. This population also begins to demonstrate the bimodality in [$\alpha$/Fe] between the discs, with some members of the young thick disc showing similar chemical properties to the older population. Only a small number of these stars have been observed, though. The significance of the population will only be known once a larger sample has been studied. The implications of the findings may be significant for understanding some of the mechanisms of Galactic evolution.

The young, $\alpha$-rich stars are not the only young stars at high $|Z|$ (e.g. \citealt{2016A&A...595A..60J,2018ApJ...852...49H}). There are many questions surrounding the young populations far from the Galactic mid-plane. These include whether these stars have formed in situ; they migrated and have been captured as they pass through this region of the Milky Way; or are they products of stellar binary evolution and mergers \citep{Izzard2018}? Each scenario has implications on the evolution of the Galactic structure. A thorough treatment of the kinematics is required to further disentangle the origins of these stars. Though not a pressing concern for this work, future exploration with \textit{Gaia} data to examine the kinematics and orbits of these stars will be undertaken to determine if they indeed belong to this population or are migrators.

The shape of the age distributions are a reflection of the star formation history of the Galaxy. A feature in common for all of the age distributions shown in Fig. \ref{fig:AgeKDE} is a minimum. The presence of this feature is indicative of a quenching period, in which the star formation rate drops and reduction in stars of certain ages is observed. With well defined peaks in both the K2 SM and APOKASC distributions, there is evidence to support two epochs of star formation. Tighter age constraints are required before the lengths of any epochs of suppressed/enhanced star formation can be confirmed, but the results are concurrent with predictions of a cessation in star formation at $\sim 9$ Gyr (e.g. \citealt{2016A&A...589A..66H}) and the predictions of the two-infall model \citep{1997ApJ...477..765C,2001ApJ...554.1044C,2009IAUS..254..191C} of distinct formation epochs for the thin ($|Z| \leq 1.0$ kpc, [$\alpha$/Fe] $\leq 0.1$, age $\sim 9$ Gyr) and thick ($|Z| > 1.0$ kpc, [$\alpha$/Fe] $> 0.1$, age $> 10$ Gyr) discs. 

\section{Conclusion} \label{conclusion}

K2 campaign fields 3 and 6 have been used to demonstrate further the capacity of K2 as a Galactic archaeology mission. The existence of observations towards the Galactic poles in addition to the \textit{Kepler} field highlight the importance of these observations to contributing towards some of the key questions surrounding the existence of the Galactic thin and thick discs. 

Initial investigations of the K2 campaigns C3 and C6, and \textit{Kepler} populations provided evidence of the K2 fields presented containing a larger proportion of metal poor stars than found with \textit{Kepler}. Further differences were found in the sample parameter distributions:

\begin{itemize}
\item \textit{Radii - } Both populations exhibited the expected prominence due to the red clump, but the K2 distribution showed an additional, more pronounced compared to \textit{Kepler}, peak at $\sim 8\, \rm{R}_{\odot}$. The feature was also found to be significant in a sample of \textit{Kepler} $\alpha$-rich stars. It relates to the RGB-bump, and its prominence in 3 fields of different pointing means it could be used as a potential indicator of $\alpha$-rich stellar populations in future. 

A shift in the position of the red clump peak was observed between the \textit{Kepler} full (APOKASC) and $\alpha$-rich (APOKASC $\alpha$-rich) samples. The $\alpha$-rich sample showed a peak 0.5 R$_{\odot}$ lower for the clump compared to the full sample. Using a TRILEGAL simulation of the \textit{Kepler} field, a divergence in the clump location was evidenced when using the simulation's Galactic component identifier. The $\alpha$-rich peak aligned with TRILEGAL expectations of the thick disc population as it is expected to be older and metal poor. The full sample aligned with the thin disc peak. The position of the peak of the red clump could therefore be considered as an indicator of the type of population being observed, but additional research beyond the scope of this paper would be required to establish this.

\item \textit{Mass - } We demonstrate the robustness of our results to population variance and underlying physical prescriptions when considering the mass against $Z$ distribution of the K2 fields. A trend of decreasing mass with increasing $|Z|$ is observed, with some scatter at high mass. We also find evidence of a decreasing metallicity gradient with decreasing mass and increasing $|Z|$. These trends remain consistent when different mass sources are used, the red clump is removed from the population and a different underlying grid is used for the analysis. The robust nature of these global trends lends confidence to our inferences of global population properties being representative of the sample.

\item \textit{Ages - }As an excellent example of one of the most metal-poor asteroseismic populations observed, the resultant K2 age distributions reveal possible epochs of enhanced star formation. The observed sample also shows the evolution of the age distribution as a function of height above the Galactic mid-plane, matching the \textit{Kepler} sample closely out to $\sim$ 1 kpc in concordance with the works by \cite{2016MNRAS.455..987C} and \cite{2018MNRAS.475.5487S}, but differing significantly beyond this distance. Further investigation clearly indicated that as $Z$ increases, the population becomes dominated by older stars. The changing distributions not only lend support to the theories indicating the thick disc is older than the thin disc, but reaffirm the desire for precise ages to allow for the confirmation of any possible epoch of quiescence in star formation and age bimodality associated with chemistry. It must be noted that some young stars ($< 4$ Gyr; possible binary products or migrators) remain at high $Z$ and require further investigation as to their origin and nature.

A strong bimodality was also observed in the age distribution for both \textit{Kepler} and the K2 fields (Fig. \ref{fig:AgeKDE}), with distinct young and old population peaks at 5 and 14 Gyr. Clear associations with $|Z|$ and [$\alpha$/Fe] were attributed to each peak: 5 Gyr - low-$\alpha$, $|Z| \leq 1.0$ kpc (thin disc); 14 Gyr - high-$\alpha$, $|Z| > 1.0$ kpc (thick disc). The chemical age dichotomy was also confirmed with the \textit{Kepler} sample, where the peak at 12 Gyr is due to the $\alpha$-rich population. Each sample presented contains a minimum, suggestive of a time delay between the formation of these populations. Given the geometric and chemical characteristics of the old and young populations, the argument can be made that this is representative of a time difference in the formation histories of the thin and thick discs.

\end{itemize}

Further work remains prior to making definitive conclusions on the true age profile of the Galactic disc. The K2$_{\rm{HQ}}$ sample shows a glimpse of what is achievable with improved precision in age determination, as does the K2 Spec. sample. However, an increase in the sample size is necessary to lend further weight to our conclusions regarding the population properties. The asteroseismic yield of stars towards the Galactic poles is continuing to increase with all sky observations from the NASA TESS mission \citep{2015JATIS...1a4003R} currently in progress. 

An increase in spectroscopic coverage of the K2 campaign fields would yield improvements in parameter determination though, with the known high quality of the asteroseismic inferences linked closely to the quality of input spectroscopic parameters. With more wide field spectroscopic surveys coming online in the near future (e.g. WEAVE, 4MOST), in addition to increased asteroseismic information from additional K2 campaign fields with observations of the Galactic poles, the field is well positioned to continue to exploit the opportunities afforded by the K2 mission to understand the true nature of the vertical structures of the Milky Way.

\section*{Acknowledgements}

We gratefully acknowledge the support of the UK Science and Technology Facilities Council (STFC).  BMR, AM, GRD, BM, LG and SK are grateful to the International Space Science Institute (ISSI) for support provided to the \mbox{asteroSTEP} ISSI International Team. AM acknowledges support from the ERC Consolidator Grant funding scheme ({\em project ASTEROCHRONOMETRY}, G.A. n. 772293). CC acknowledges support from DFG Grant CH1188/2-1 and from the ChETEC COST Action (CA16117), supported by COST (European Cooperation in Science and Technology). BMR would like to thank the AIP for temporarily hosting me during the studies for this work. 
LC is the recipient of the ARC Future Fellowship FT160100402. SM acknowledges support from NASA grants NNX16AJ17G and NNX15AF13G, by the National Science Foundation grant AST-1411685 and the Ramon y Cajal fellowship number RYC-2015-17697. RAG acknowledges the funding revived from the CNES through the PLATO grant. PJ acknowledges FONDECYT Iniciaci\'on Grant Number 11170174. TSR acknowledges financial support from Premiale 2015 MITiC (PI B. Garilli). AG acknowledges support from the Swedish National Space Board.
Funding for the Stellar Astrophysics Centre is provided by the Danish National Research Foundation (Grant DNRF106). 

Funding for the Sloan Digital Sky Survey IV has been provided by the Alfred P. Sloan Foundation, the U.S. Department of Energy Office of Science, and the Participating Institutions. SDSS acknowledges support and resources from the Center for High-Performance Computing at the University of Utah. The SDSS web site is www.sdss.org.

SDSS is managed by the Astrophysical Research Consortium for the Participating Institutions of the SDSS Collaboration including the Brazilian Participation Group, the Carnegie Institution for Science, Carnegie Mellon University, the Chilean Participation Group, the French Participation Group, Harvard-Smithsonian Center for Astrophysics, Instituto de Astrofisica de Canarias, The Johns Hopkins University, Kavli Institute for the Physics and Mathematics of the Universe (IPMU) / University of Tokyo, the Korean Participation Group, Lawrence Berkeley National Laboratory, Leibniz Institut f\"ur Astrophysik Potsdam (AIP), Max-Planck-Institut f\"ur Astronomie (MPIA Heidelberg), Max-Planck-Institut f\"ur Astrophysik (MPA Garching), Max-Planck-Institut f\"ur Extraterrestrische Physik (MPE), National Astronomical Observatories of China, New Mexico State University, New York University, University of Notre Dame, Observat\'orio Nacional / MCTI, The Ohio State University, Pennsylvania State University, Shanghai Astronomical Observatory, United Kingdom Participation Group, Universidad Nacional Aut\'onoma de M\'exico, University of Arizona, University of Colorado Boulder, University of Oxford, University of Portsmouth, University of Utah, University of Virginia, University of Washington, University of Wisconsin, Vanderbilt University, and Yale University.

% ...
\bibliographystyle{mnras}
\bibliography{main}




\end{document}